\documentclass[useAMS,usenatbib]{mn2e}

\usepackage[pdfpagemode={UseOutlines},bookmarks,bookmarksopen,colorlinks,citecolor={blue},urlcolor={red}]{hyperref}
\usepackage[flushleft]{threeparttable}
\usepackage{booktabs,caption}
\usepackage{graphicx}
\usepackage{amsmath}
\usepackage{amssymb}
\usepackage{tabularx}
\usepackage{float}
\usepackage{etoolbox}
\usepackage{color}
\usepackage{natbib}

\newcommand{\erf}{\mathrm{erf}\,}
\newcommand{\Ms}{~{\rm M}_\odot}
\newcommand{\bhb}{{\rm BHB}}
\newcommand{\gw}{{\rm GW}}
\newcommand{\bh}{{\rm SMBH}}
\newcommand{\gc}{{\rm GC}}
\newcommand{\ibh}{{\rm IMBH}}

\newcommand{\HiGPUs}{\texttt{HiGPUs}~}
\newcommand{\ARGdf}{\texttt{ARGdf}~}
\newcommand{\ARCHAIN}{\texttt{ARCHAIN}~}

\makeatletter

\patchcmd{\NAT@citex}
  {\@citea\NAT@hyper@{%
     \NAT@nmfmt{\NAT@nm}%
     \hyper@natlinkbreak{\NAT@aysep\NAT@spacechar}{\@citeb\@extra@b@citeb}%
     \NAT@date}}
  {\@citea\NAT@nmfmt{\NAT@nm}%
   \NAT@aysep\NAT@spacechar\NAT@hyper@{\NAT@date}}{}{}

\patchcmd{\NAT@citex}
  {\@citea\NAT@hyper@{%
     \NAT@nmfmt{\NAT@nm}%
     \hyper@natlinkbreak{\NAT@spacechar\NAT@@open\if*#1*\else#1\NAT@spacechar\fi}%
       {\@citeb\@extra@b@citeb}%
     \NAT@date}}
  {\@citea\NAT@nmfmt{\NAT@nm}%
   \NAT@spacechar\NAT@@open\if*#1*\else#1\NAT@spacechar\fi\NAT@hyper@{\NAT@date}}
  {}{}

\makeatother

\title[The MEGaN project II]{The MEGaN project II. Gravitational waves from intermediate mass- and binary black holes around a supermassive black hole}
\author[Arca-Sedda, M. and Capuzzo-Dolcetta, R.]{M. Arca-Sedda$^{1}$ \thanks{E-mail: m.arcasedda@gmail.com}, R. Capuzzo-Dolcetta$^2$\\
$^{1}$Zentrum f\"{u}r Astronomie der Universit\"{a}t Heidelberg, Astronomisches Rechen-Institut, M\"{o}nchhofstr. 12-14, D-69120, Heidelberg (Germany)\\
$^{2}$Dept. of Physics, Sapienza, University of Rome, Piazzale Aldo Moro 5, I-00185, Rome (Italy)\\
}

\begin{document}
\date{Revised to }

\pagerange{\pageref{firstpage}--\pageref{lastpage}} \pubyear{2015}

\maketitle

\label{firstpage}

\maketitle

\begin{abstract}
{ 
We investigate the evolution of intermediate-mass (IMBHs), stellar (BHs) and binary black holes (BHBs), deposited near a supermassive black hole (SMBH) by a population of massive star clusters. Stellar BHs rapidly segregate around the SMBH, driving the formation of extreme mass-ratio inspirals that coalesce at a rate $\Gamma= 0.02-0.2$ yr$^{-1}$ Gpc$^{-3}$ at redshift $z=0$. A few IMBHs orbiting the SMBH favour the formation of massive pairs that coalescence within a Hubble time, being the merger rate for this channel $\Gamma =0.03$ yr$^{-1}$ Gpc$^{-3}$. Recoiling kicks post-merger can eject the remnant from the galaxy centre, especially in dwarf galaxies. Our results suggest that this mechanism can lead to up to $10^5$ ejected SMBH within 1 Gpc. An IMBH co-existing with a few single and binary BHs in the same cluster can affect significantly their evolution, either driving binary disruption, yielding to intermediate-mass ratio inspirals (merger rate $\Gamma =9.5$ yr$^{-1}$ Gpc$^{-3}$), or boosting BHBs coalescence ($\Gamma =2-8$ yr$^{-1}$ Gpc$^{-3}$). In a few simulations, the SMBH boosts BHBs coalescence, leading this process to a merger rate $\Gamma =1$ yr$^{-1}$ Gpc$^{-3}$. We note that BHBs experiencing a merger in a galactic nucleus can be erroneously estimated $\sim 30\%$ heavier than it really is because of the Doppler shift of the wave frequency as caused by the rapid motion around the SMBH. All our simulations are carried out using an $N$-body code tailored to treat close encounters and post-Newtonian dynamics, that includes also the galaxy field and dynamical friction in the particles equation of motion.
}
\end{abstract}


\section{Introduction}

Globular clusters (GCs) are thought to be the birthplace of intermediate mass black holes (IMBHs), an elusive class of objects with masses in the range $\sim 10^2-10^5\Ms$ that should fill the gap between stellar (BHs) and supermassive black holes (SMBHs).
Probing the existence of IMBHs is one of the major challenges of modern astrophysics, due to the small observational fingerprints that they leave on their surroundings \citep{noyola10, vandermarel10, haggard13, lanzoni13, Lutzgendorf13, Lutz15, kiziltan17}. 
 
One possible scenario for IMBH formation is via repeated stellar collisions in dense and massive GCs (runaway scenario), according to which the IMBH buildup occurs over the host GC core-collapse time-scale, i.e. $1-10$ Gyr \citep{zwart02,zwart07,Giersz15}.
As recently discussed by \cite{Giersz15}, runaway collisions can drive the IMBH growth either through a {\it slow} or {\it fast} mechanism, depending on the GC structure. The {\it slow} mechanism requires that all the BHs forming inside the GC core are ejected but one. The remaining BH starts growing through stellar consumption slowly, reaching the IMBH mass range ($\gtrsim 150\Ms$) over relatively long time-scales $>1$ Gyr. 
Conversely, the {\it fast} mechanism shows that an extreme GC central density, $\sim 10^8\Ms$ pc$^{-3}$, can drive the prompt IMBH formation through multiple BH-BH collisions. However, whether such large densities can be reached during the GC formation process is still unclear. 

\cite{Giersz15} results indicate that the probability for IMBH formation through both the channels is around $20\%$, with only $5\%$ of IMBHs forming through the fast mechanism. 

Inefficient BH-BH and BH-stars interactions can lead to the formation of a long-living BH subsystem \citep{breen13,AS16,AAG18a,AAG18b}, being unable to eject most of the BHs, which is required by the slow mechanism, or make them merge promptly, according to the fast mechanism.
Moreover, BH-BH mergers may result in the ejection of the merger product via gravitational waves (GW) recoil, thus decreasing the probability for an IMBH seed to grow through subsequent mergers or stellar consumption.

Aside from the runaway scenario, other IMBH formation mechanisms are widely debated in the recent literature. For instance, massive Pop III stars, which are expected to form in the early Universe with an extremely low metal content, are thought to undergo direct collapse and form IMBHs with masses as high as $10^2-10^4\Ms$ \citep{madau01,schneider02,bromm02,bromm03,ohkubo09,spera17}. Moreover, IMBHs with masses $\sim 10^3\Ms$ can form via direct collapse of gaseous clouds in the primordial Universe, as suggested by \cite{latif13}, or might form in metal poor galactic halos at high redshift \citep{bellovary11}. Another possibility is that IMBHs might form in the centre of satellite galaxies, which are stripped away as they merge with the main galaxy. According to this scenario, galactic halos similar to the Milky Way could potentially host a population of $5-15$ wandering IMBHs \citep{bellovary10}.
Also, IMBHs can form already in the galactic nuclei, either in AGN accretion discs \citep{mckernan12}, or in circumnuclear regions of disc galaxies \citep{taniguchi00}.
 
GCs are also the perfect nurseries for the formation of stellar BH binaries (BHBs). The recent detection of gravitational waves (GWs) operated by the LIGO-Virgo collaboration \citep{abbott16a,abbott16b,abbott16c,abbott16e,abbott17a,abbott17b,abbott17c,abbott17d,abbott17e}, opened new chapters in astronomy and astrophysics. 
The LIGO detections spotted for the first time the merging of two black holes (BHs) with masses $20 \Ms$ and $30\Ms$, providing the evidence for BHs with masses above $20\Ms$, as well as of stellar black hole binaries \citep{abbott16d}. The subsequent discovery of new mergers events by BHBs with a mass in between $15\Ms$ and $50\Ms$ \citep{abbott16b,abbott16c,abbott17a,abbott17b} and the first coalescence of a neutron star binary \citep{abbott17c,abbott17d}, which started the ``multi-messenger'' astronomy era, pulled the scientific community in spending a large effort to unveil the origin of GWs sources in different environments \citep{rodriguez15,askar17,antonini16,belczinski17,ASG17}.

Merging BHBs can form in massive GCs \citep{rodriguez15,rodriguez16,askar17}, young massive clusters \citep{mapelli16,banerjee16}, in dwarf spheroidal galaxies in the local volume \citep{schneider17},  or even via strong dynamical interactions in galactic nuclei \citep{bartos16,antonini16b,ASG17,hoang18}.
As opposed to this ``dynamical scenario'', merging BHBs can also form via isolated binary evolution \citep{belczynski16}, or through the axisymmetric fragmentation of a rapidly rotating massive star into two heavy BHs \citep{loeb16}.

Although different BHBs merging channels might be characterized by different properties, it is currently unclear whether the detected population of BHB mergers can provide information on the environment in which they formed \citep{seoane16,ASBE18}.

The possible formation and evolution of IMBHs and BHBs in GCs can have interesting implications for the evolution of the host galaxy nucleus. Indeed, GCs can delivery their IMBHs and/or BHBs as they sink toward the galactic centre due to dynamical friction.   
If the galaxy hosts an SMBH, the resulting formation of complex interactions with the delivered IMBHs and the BHBs can potentially lead to several phenomena, like BHB mergers via Lidov-Kozai cycles \citep[see for instance][]{antonini12}, formation of extreme- (EMRIs) and intermediate-mass ratio inspirals (IMRIs) \citep{seoane07,amaro12,seoane15}, or the formation of IMBH-SMBH pairs \citep{miller05}. 

The GWs echoes produced by such systems can be potentially detected by current and future GW observatories, although their formation probability in galactic nuclei is still uncertain.

In this paper, we investigate how a population of stellar BHBs and IMBHs, delivered toward the centre of a galaxy with mass $10^{11}\Ms$ by a population of massive GCs, interacts with an SMBH ($M_\bh=10^8\Ms$) inhabiting the galactic centre.

We combine the results obtained through a direct N-body simulation of the galaxy and the GCs population, called MEGaN, with detailed few-body integrations that model the complex interactions between several IMBHs or BHBs orbiting around the galaxy SMBH.

Few-body simulations are performed using \ARGdf, a direct N-body code suited to represent few-body dynamics, which include post-newtonian formalism, the effect of an external potential and dynamical friction. The code represents an extension of the \ARCHAIN code \citep{mikkola99,mikkola08}, developed to model strong gravitational interactions through the so-called algorithmic regularization method \citep{mikkola99}.

The paper is organised as follows: in Section \ref{nsc} we discuss the implications of GCs orbital segregation on the evolution of galactic nuclei, in Section \ref{model} we briefly describe the galaxy model and the GC orbital and structural properties of the MEGaN simulation, while in Sect. \ref{EMRI} we use the results of the model to calculate the formation rate of EMRIs; in Section \ref{ARGdf} we present \ARGdf, providing details on the procedure used to treat dynamical friction in few-body dynamics; Section \ref{results} is devoted to discuss all the channels investigated through \ARGdf in the paper, i.e. SMBH-IMBH, IMBH-BH, IMBH-BHB and SMBH-BHB pairing and evolution.

\section{Delivering IMBHs and BHBs in galactic nuclei}
\label{nsc}

For the sake of clarity, hereby we refer to massive clusters as GCs, independently of their age, mass or metallicity.
While BHB mergers are frequent in an early phase of the host GC life-time, the formation of an IMBH is a slower process, and, possibly, shapes significantly the host cluster properties \citep{Giersz15,zocchi}.
Most of the studies presented in literature investigate the formation and evolution of BHBs and IMBHs supposing GCs orbiting very far from the centre of their galaxy host, where the galactic gravitational field has a marginal effect on the star cluster evolution.

On the other hand, if a massive cluster forms in the inner $0.5-1$ kpc portion of its galaxy, its motion is substantially altered by two large scale processes: dynamical friction, which drags the cluster toward the galactic centre, and tidal forces, which strip away its stars \citep{ernst09}.
The competing action of dynamical friction and galactic tidal forces are thought to be at the base of nuclear star cluster (NSC) formation \citep{Trem75, TrOsSp, Trem76, Dolc93, AMB, antonini13, perets14, ASCD14b, gnedin14, ASCD15He, ASCDS16, ASCD16b}.

NSCs are massive and compact stellar systems, with masses in the $10^6-10^8\Ms$ mass range and half-light radii $1-5$ pc, observed in the centre of a large fraction of galaxies belonging to different mass classes and Hubble types \citep{cote06,Turetal12}. NSCs are characterised by a complex star formation history, possibly connected with their origin \citep{rossa,walcher06,carson15}. Often, they harbour a central SMBH, with a mass $M_{\rm SMBH}\leq 10^8\Ms$  \citep{graham12,Neum,georgiev16,CDTEM17}.
The Milky Way is not an exception, it hosts a NSC with mass $2.5\times 10^7\Ms$ and effective radius $\sim 2$ pc \citep{schodel14,gallegocano2017,schodel2017}, harbouring an SMBH with mass $3.6^{+0.2}_{-0.4}\times 10^6 \Ms$ \citep{schodel02,ghez08,gillessen09,schodel09}.

The Galactic NSC is characterised by two distinct stellar populations: a young component, mostly concentrated in the NSC inner region and likely formed in-situ $\sim 3-8$ Myr ago, and an older population with an average age $\simeq 10$ Gyr \citep{lu09,bartko09,do13,lu13,pfuhl11,feldmeier15,min16}.
In the dry-merger scenario, during the NSC formation GCs undergo a series of strong encounters with the central SMBH, if present. These interactions can strongly affect the shape and properties of the galactic central region, \citep{perets07,antonini14, aharon16, ASCDS16, ASCD17}. 

In this context, the effects of the interaction between the SMBH and an infalling GC on BHs population (either stellar or intermediate-mass) has not yet been quantitatively investigated due to the heavy computational load required to investigate deeply on such a problem.

If several GCs forms well deep in the galaxy innermost regions, their orbital decay and subsequent interaction with the central SMBH will occur before the BHs and BHB population had time to evolve. Hence, a number of BHs and BHBs can be deposited around the SMBH, where the intense tidal forces might shape their overall evolution.

If the GCs formed in an outer region of the galaxy, instead, their segregation time-scale can easily exceed several Gyr, allowing the GCs to possibly be the site of IMBH formation. The subsequent orbital decay of multiple GCs can bring together several IMBHs around an SMBH \citep{mastrobuono14,ASG17,fragione17c,baumgardt06}. 
\cite{mastrobuono14} outlined that a population of IMBHs, dragged to the MW NSC during its formation, should have left evident fingerprints in the kinematics of stars moving in the inner few pc of the MW which, however, are not observed.
One possibility is that IMBHs are kicked out from the parent cluster during their formation, as a consequence of dynamical kicks or GW recoiling kicks developed during an IMBH-BH merger events \citep{bockelmann08,konstantinidis13,fragione17c}, although this seems to be unlikely in the so-called \textit{slow scenario} for IMBH formation \citep{Giersz15}. Using a huge set of massive star clusters models, 
\cite{Giersz15} demonstrated that the nature of IMBH formation is a highly `stochastic' phenomenon, strongly dependent on the host cluster properties. 

All the considerations above indicate that the presence of many IMBHs surrounding an SMBH is unlikely, being much more probable to have there just a few of them.

In this paper, we use the outcomes of a direct $N$-body model representing the nucleus of a massive elliptical galaxy to investigate the evolution of IMBHs and BHBs in galactic nuclei. 
Our numerical model, called ``MEGaN'', simulate the evolution of 42 massive GCs orbiting an SMBH with mass $10^8\Ms$ \citep{ASCD17}.
Using the detailed information on the dynamics of the GCs and the galactic nucleus obtained through our MEGaN simulation, we develop several specific models (as we will detail later on) tailored to model two possible scenarios:
\begin{itemize}
\item the GC population formed in an inner portion of the galaxy. In this case, the GC - SMBH gravitational interactions occur when GCs are still in an early phase of their life;
\item the GC population formed outside the galaxy bulge, and slowly spiralled toward the inner regions over $\sim 1-10$ Gyr. In this case, most of the processes related to GC internal evolution should have already taken place, leading to the possible formation of an IMBH. 
\end{itemize}

Currently, it is hard to determine whether a NSC formed on a long time-scale from GCs coming far out from the galactic centre, or on a much shorter time from GCs born well inside the galactic inner bulge. 
Likely, in the dry merger scenario, NSC formation arises from a combination of both the cases: in a first phase clusters born closer to the galactic center rapidly segregate and merge, leading to the formation of an ``NSC seed'';  other clusters infall later and merge into this growing-up NSC. Hopefully, observations at high-resolution of the $\gamma$ and X-ray flux coming from the MW NSC and their interpretation will shed light on which is the dominant phase in the NSC build-up \citep{hooper11,perez15,brandt15,ASK17,fermi17,fragione17b,abbate18}.

In this framework, our paper investigates the consequences of strong scattering interactions between infalling GCs and a central SMBH in a massive elliptical galaxy. 
In particular, we point the attention on three different processes. 
In first place, assuming that some GCs formed outside the galaxy core and reached the inner region on a relatively long time ($> 1$ Gyr), we investigate the interactions between 2 IMBHs, dragged to the galactic centre by their hosting most massive and densest GCs, and the central SMBH. Our results suggest a high probability to observe an IMBH-SMBH collisions within on time-scales of $10^2$ Myr,
partially explaining the dearth of IMBHs kinematic signatures in galactic nuclei. 

On another hand, we explore also the possibility that at least one BHB orbits the IMBH, aiming at determining whether or not the BHB can survive long enough to be transported closer to the galactic SMBH. 

Finally, in the case in which the IMBH formation fails, we study the possibility that several single BHs or BHBs are left orbiting around the SMBH in consequence of the GC orbital decay and disruption. 
In this regard, two are the possible outcomes: i) some BHs are captured on very tight orbits, leading to an EMRI \citep{hils95,seoane07,merritt11,amaro12,seoane15}, or ii) the coalescence of BHBs orbiting the SMBH is boosted by the Kozai-Lidov mechanism \citep{lidov62,kozai62}.

EMRIs (mass ratio above $10^4$) and IMRIs \citep{seoane07}, are one of the most promising sources of GWs to be detected with the next generation of space-based GW observatories, Laser Interferometer Space Antenna\footnote{\url{https://www.lisamission.org/}} \citep[LISA,][]{mapelli12,eLISA13,vitale14} and the Chinese space interferometers ``TianQin'' \citep{tianqin16} and ``Taiji'' \citep{taiji17}. Thus, characterising their occurrence in galactic nuclei represents a key to correctly interpret possible future observations.

\section{Direct N-body modelling of a massive elliptical galaxy nucleus}
\label{model}

In the first paper of this series, we introduced the MEGaN numerical simulation, which represents one of the first simulations of an entire massive galactic nucleus hosting an 
SMBH with mass $10^8\Ms$ and 42 GCs with masses in the range $(0.3-2)\times 10^6\Ms$ \citep{ASCD17}. 
The MEGaN model was evolved through the \texttt{HiGPUs} code \citep{Spera}, a direct summation, 6$^{\rm th}$ order Hermite integrator with block time-steps, which runs efficiently on composite GPU+CPU platforms. 
As deeply discussed in the paper, the combined action of the SMBH and the galactic background affects significantly the GC evolution, yielding to their disruption in several cases. On another hand, a non-negligible amount of GCs debris can accumulate around the SMBH possibly leading, for instance, to an enhancement of tidal disruption events.

In this section, we will make use of the simulation outcomes to explore the possibility that the GC disruption may affect the possible formation of gravitational wave sources around the SMBH.

\subsection{The galaxy model and its globular cluster system}

The MEGaN galaxy model is represented by a truncated Dehnen's density profile \citep{Deh93}:
\begin{equation}
\rho_{\rm D}(r)=\frac{(3-\gamma)M_g}{4\pi r_g^3}\left(\frac{r}{r_g}\right)^{-\gamma}\left(1+\frac{r}{r_g}\right)^{-4+\gamma}\frac{1}{{\rm cosh}(r/r_{\rm cut})} ,
\label{dens}
\end{equation}
where $M_g = 10^{11}\Ms$ is the total galaxy mass, $r_g = 2$ kpc its scale radius, $\gamma = 0.1$, and $r_{\rm cut}$ is a {\it truncation} radius. The choice $r_{\rm cut} = 70$ pc allows us to represent both the galaxy nucleus and each of the 42 GCs by a suitable sample of particles. To model the whole (SMBH + galaxy + GC) system we used $2^{20}\gtrsim 10^6$ particles, thus representing one of the most refined N-body models of a galactic nucleus to date.

All the GCs were represented through \cite{king62} models, core radii spanning the range $0.2 - 1.4$ pc, initial apocentres in between $30-150$ pc and eccentricities according to a thermal distribution \citep{jeans19}.  
As discussed in \citet{ASCD17}, this choice of parameters allowed us to provide a reliable galaxy environment for studying the GC orbital evolution. 

The main GC properties are described in table \ref{GCS}. 
We refer the reader to the main paper for further information about the GC initial conditions and late evolution.

\begin{table*}
\caption{}
\centering{Properties of the GCs sample}
\begin{center}
\begin{tabular}{lccccccccc}
\hline
\hline
\multicolumn{1}{c}{GC}  & $W_0$ & $R_t$ & $R_c$ & $M_{\rm GC}$ & $r_{\rm GC}$ & $v_{\rm GC}$ & $e$ & $t_{\rm df}$ & $N_{\rm GC}$\\
\multicolumn{1}{c}{name} &       & (pc)  & (pc)  & $10^6$ M$_\odot$ & (pc) & km s$^{-1}$ &  & (Gyr) & \\ 
\hline
GC1 &7.54 &10.9 &0.207 &1.05 &71.4 & 121 & 0 & 0.295 &  10445 \\
GC2 &7.13 &17 &0.443 &0.906 &117 & 89.4 & 0.527 & 0.509 &  8995 \\
GC3 &7.66 &23.8 &0.424 &1.68 &134 & 87.7 & 0.483 & 0.446 &  16671 \\
GC4 &7.08 &13.7 &0.378 &1.89 &74.3 & 115 & 0.974 & 0.0755 &  18754 \\
GC5 &7.89 &16.5 &0.257 &1.1 &107 & 91.6 & 0.553 & 0.368 &  10966 \\
GC6 &7.28 &15.1 &0.35 &0.452 &132 & 24.5 & 0.884 & 0.633 &  4492 \\
GC7 &7.71 &23.2 &0.397 &1.69 &130 & 93.5 & 0.687 & 0.336 &  16792 \\
GC8 &7.78 &20.8 &0.345 &1.07 &136 & 78.2 & 0.177 & 0.804 &  10634 \\
GC9 &6.88 &15.2 &0.477 &1.87 &82.6 & 91.3 & 0.32 & 0.204 & 18554 \\
GC10 &6.58 &22 &0.812 &1.16 &140 & 33.6 & 0.785 & 0.432 &  11560 \\
GC11 &7.33 &22.6 &0.502 &1.25 &140 & 86.2 & 0.411 & 0.629 & 12412 \\
GC12 &7.49 &16.2 &0.32 &1.62 &92.6 & 93 & 0.478 & 0.239 &  16054 \\
GC13 &7.27 &14.9 &0.347 &0.85 &105 & 78 & 0.129 & 0.618 &  8446 \\
GC14 &6.54 &15.4 &0.58 &0.719 &115 & 76.9 & 0.125 & 0.806 &  7137 \\
GC15 &6.76 &25.1 &0.839 &1.66 &142 & 61.4 & 0.29 & 0.586 & 16530 \\
GC16 &7.74 &12.7 &0.214 &1.28 &78.4 & 121 & 0 & 0.305 & 12716 \\
GC17 &7.01 &8.78 &0.257 &1.86 &47.7 & 127 & 0.729 & 0.051 &  18517 \\
GC18 &6.5 &5.01 &0.194 &0.583 &40.1 & 131 & 0.561 & 0.0997 & 5791 \\
GC19 &6.45 &16.6 &0.665 &0.834 &118 & 69.8 & 0.0695 & 0.801 &  8284 \\
GC20 &6.11 &14.4 &0.727 &0.33 &139 & 85 & 0.38 & 1.56 &  3273 \\
GC21 &7.45 &13.9 &0.283 &1.49 &81.5 & 90.9 & 0.31 & 0.234 &  14840 \\
GC22 &7.1 &19.7 &0.53 &1.38 &119 & 72.6 & 0.00601 & 0.599 & 13721 \\
GC23 &7.14 &19.1 &0.497 &1.7 &107 & 93.5 & 0.616 & 0.26 &  16836 \\
GC24 &6.96 &11.6 &0.352 &1.3 &71.5 & 122 & 0 & 0.257 & 12871 \\
GC25 &6.63 &16 &0.574 &0.613 &126 & 83.1 & 0.32 & 0.903 &  6090 \\
GC26 &6.66 &26.2 &0.921 &1.97 &140 & 55.8 & 0.407 & 0.462 &  19521 \\
GC27 &7.78 &19.6 &0.324 &1.24 &122 & 36.7 & 0.743 & 0.343 & 12353 \\
GC28 &7.88 &8.85 &0.138 &1.27 &54.7 & 136 & 0 & 0.163 &  12598 \\
GC29 &6.02 &26.8 &1.46 &1.79 &148 & 93.3 & 0.641 & 0.426 &  17756 \\
GC30 &7.2 &18.5 &0.456 &1.03 &122 & 93.9 & 0.686 & 0.421 &  10226 \\
GC31 &6.43 &14.4 &0.583 &0.612 &114 & 56.6 & 0.392 & 0.711 &  6073 \\
GC32 &7.57 &15.1 &0.284 &0.845 &107 & 77.5 & 0.111 & 0.647 &  8397 \\
GC33 &6.71 &9.43 &0.323 &1.03 &62.5 & 135 & 0 & 0.236 & 10258 \\
GC34 &7.17 &19.1 &0.484 &1.08 &124 & 100 & 0.92 & 0.298 &  10776 \\
GC35 &7.25 &11.5 &0.272 &0.486 &98 & 62.1 & 0.316 & 0.683 &  4825 \\
GC36 &7.09 &19.7 &0.536 &1.25 &122 & 34 & 0.778 & 0.328 &  12369 \\
GC37 &6.47 &24.3 &0.957 &1.69 &137 & 79.6 & 0.217 & 0.576 &  16787 \\
GC38 &7.6 &14.5 &0.268 &0.334 &140 & 85.2 & 0.379 & 1.56 &  3318 \\
GC39 &6.16 &23.9 &1.17 &1.41 &143 & 84.3 & 0.336 & 0.636 &  13990 \\
GC40 &7.18 &23 &0.575 &1.56 &133 & 87.6 & 0.485 & 0.458 & 15468 \\
GC41 &7.74 &23.7 &0.401 &1.39 &142 & 71.1 & 0.0492 & 0.8 &  13768 \\
GC42 &7.2 &4.55 &0.112 &0.478 &38.9 & 141 & 0.781 & 0.0828 & 4747 \\
\hline
\end{tabular}
\end{center}
\begin{tablenotes}
\item Column 1: GC name. Column 2: value of the adimensional potential well. Column 3: GC tidal radius. Column 4: GC core radius. Columns 5-7: GC mass, initial position and velocity. Column 8: GC orbital eccentricity. Column 9: dynamical friction timescale according to Eq. \ref{tdf}. Column 10:  number of particles used to model the GCs.
\end{tablenotes}
\label{GCS}
\end{table*}

In this numerical model, the dynamical friction acting on each GC comes out naturally from its two body interactions with field stars, while tidal disruption is mainly driven by the gravitational field generated by the galaxy nucleus as a whole and the central SMBH.
As discussed in our companion paper, we stop the MEGaN simulation after $\simeq 290$ Myr, when the intense action of tidal forces have almost completely destroyed the GCs in our sample.

GCs start losing stars as long as they passes near the galactic centre. These stars are either ejected away in form of high or even hyper-velocity stars \citep{ASCDS16,fragione15}, or they can be captured by the SMBH. In the latter case, the GC disruption can enhance the rate at which stars are ripped apart from the SMBH tidal field, as shown in the companion paper \citep{ASCD17}. 

In the next section, we will explore the possibility that a fraction of these stellar debris
is comprised of compact objects that are captured by the SMBH as EMRIs.

\subsection{Extreme mass ratio inspirals}
\label{EMRI}

In our million-body simulation, the disrupting GCs leave a significant amount of stars around the SMBH, causing an evident density increase in the inner $10$ pc of the galaxy \citep[see our companion paper][]{ASCD17}. 
A steeper density profile surrounding the SMBH can result in a larger probability for stars to either fall onto the SMBH, the so-called \textit{direct plunge}, or bind to the SMBH on a tight orbit and form an EMRI \citep{hils95,seoane07,merritt11}.

In the case of Schwarzschild's SMBHs, an EMRI event requires that the time over which the star orbit changes due to GW emission is shorter than the 2-body relaxation time-scale \citep[see][for a detailed review on this process]{seoane07}.

This condition can be written as 
\begin{equation}
t_\gw < (1 - e)t_{\rm rel}.
\end{equation}

Manipulating the two sides of the inequality allows finding the threshold eccentricity required for an EMRI to form
\begin{equation}
e_t < 1 - \frac{t_\gw}{t_{\rm rel}}.
\end{equation}

Given the eccentricity lower limit for directly plunging orbits, $e_p = 1-4R_S/a$, where $R_S$ is the SMBH Schwarzschild radius and $a$ is the EMRI semi-major axis, clearly an EMRI can develop only if $e_t<e_p$. 

This yields to definition of a critical semi-major axis $a_{\rm EMRI}$ as such below which the star can be captured by the SMBH as an EMRI, given by

\begin{align}
a_{\rm EMRI} (pc) = & 5.3\times 10^{-2} \left(\frac{t_{\rm rel}}{1 {\rm Gyr}}\right)^{2/3} \times\\\nonumber
  & \times 
 \left(\frac{m_*}{10\Ms}\right)^{2/3}
 \left(\frac{M_\bh}{10^6\Ms}\right)^{-1/3}, 
\label{emrisemi}
\end{align}
so that heavier stars are most likely captured as EMRIs. 

As discussed in the companion paper \citep{ASCD17}, the stellar pericentre distribution in the MEGaN simulation is well described by a simple relation, $f(r_p)$, thus allowing calculating in an easy way the number of stars having a pericentre, $r_{\rm EMRI} = a_{\rm EMRI}(1-e)$, such to give rise to an  EMRI.

Within the SMBH influence radius, the relaxation time can be written as \cite{amaro13b}
\begin{align}
t_{\rm rel} \simeq & 0.2 {\rm ~Gyr} \left(1+\gamma\right)^{-3/2}  \times\left(\frac{\sigma}{100~{\rm km}~{\rm s}^{-1}}\right)^3  \times \nonumber \\
                   &  \times\left(\frac{10\Ms}{m_*}\right)   \times\left(\frac{10^6\Ms~{\rm pc}^{-3}}{\rho}\right),
\end{align}
where $\gamma$ is the galaxy density slope, $\sigma$ the velocity dispersion, $m_{\rm BH}$ the average mass of EMRIs candidate and $\rho$ the average stellar density within the influence radius. 

At a typical distance of 0.01 pc, our N-body model is characterized by $\sigma \simeq 120$ km s$^{-1}$ and $\rho \simeq 4300 \Ms$ pc$^{-3}$, thus implying  $t_{\rm rel} \simeq 38$ Gyr, thus larger than a Hubble time. 
However, it must be noted that a number of processes can accelerate the relaxation around the SMBH. 

Further GCs forming outside the galactic nucleus and infalling on longer time-scales can increase the density around the SMBH, thus further reducing $t_{\rm rel}$. 

Moreover, the presence of GCs and their debris in the galactic nucleus represent a simple way to make the relaxation faster. As firstly noted by \cite{spitzer1951} a number of massive perturbers like giant molecular clouds or GCs can dominate stellar scattering in the galactic nucleus, decreasing the relaxation time by a factor up to $10^2$ \citep[see also][]{Merri13,perets2007}. A further element that must be taken into account is that our galaxy model is triaxial after the GC disruption, as discussed in our companion paper \citep{ASCD17}. Triaxiality is a further element that can accelerate relaxation, due to the possibility for stars to move on orbits which pass arbitrarily close to the galactic centre. Also an axisymmetric structure can accelerate relaxation. The influence of a sub-pc disc might be at the origin, for instance, of the peculiar eccentricity distribution of the so-called S-stars in the Milky Way centre \citep{chen14}. All the above said, we assume that our galaxy nucleus has a maximum relaxation time  $t_{\rm rel} \simeq 10$ Gyr in equation \ref{emrisemi}, thus corresponding to $r_{\rm EMRI} = (1.1 - 5.3)\times 10^{-2}$ pc.

\begin{figure}
\centering
\includegraphics[width=8cm]{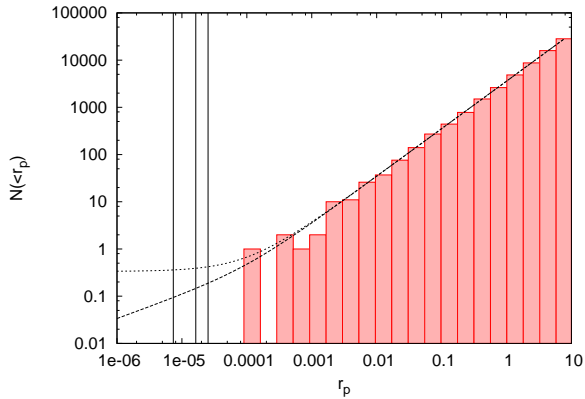}
\caption{Cumulative distribution of the stars pericentre in the MEGaN simulation. From left to right, the black vertical lines represent the Roche radius for a $0.5$, $2$ and $10 \Ms$ star. the two curves represent the fitting functions described in Eqs. \ref{f1} and \ref{f2}.
}
\label{TDEdist}
\end{figure}

Fig. \ref{TDEdist} shows the cumulative distribution, $N(<r_p)$, of the stars pericentre in our full $N$-body model. Although this simulation represents the current state-of-art for $N$-body modelling of galactic nuclei, our resolution is not sufficient to state clearly what is the behaviour of the $N(<r_p)$ low-end tail. Hence, in order to estimate the number of stars that can form an EMRI, we 
provide two different fitting functions: one producing an $N(<r_p)$ steeply decreasing to 0 and the other saturating to a nearly constant value at decreasing $r_p$. 

The rapidly decreasing function $f_1$, is defined as
\begin{equation}
f_1(r_p) = kc(ar_p+1)^{b}\sqrt{r_p},
\label{f1}
\end{equation}
while the other ($f_2$) is given by
\begin{equation}
f_2(r_p) = kc(ar_p+1)^{b}.
\label{f2}
\end{equation}

In both the equations, $k=1/N_{\rm GCS}$ represents the inverse of the number of particles used to represent all the GCs. Moreover, we set $1/a=10^{-4}$ pc, which is the length scale below which our resolution in $N(<r_p)$ loses quality. The value of the fitting parameters are resumed in table \ref{fit}.

\begin{table}
\caption{Parameters of the $N_{r_p}$ fitting functions}
\begin{center}
\begin{tabular}{cccc}
\hline
	& $a$ & $b$ & $c$\\
\hline
$f_1(r_p)$ & $10^4$ & $0.507 \pm 0.003$& $34 \pm 1$\\
$f_2(r_p)$ & $10^4$ & $1.007 \pm 0.003$& $0.33 \pm 0.01$ \\
\hline     
\end{tabular}
\end{center}
\label{fit}
\end{table}
The number of stars having orbital properties satisfying the condition for EMRIs formation is then given by:
\begin{equation}
N_{\rm EMRI}(<r_{\rm EMRI}) = f_i(r_{\rm EMRI})\frac{M_{g,sim}}{\langle m\rangle}f(m_*)f_{\rm ret}(m_*),
\label{nemris}
\end{equation}
where the weighting function $f(m_*)$ represents the fraction of stars with mass $m_*$ and $f_{\rm ret}(m_*)$ their retaining fraction, $M_{g,sim} =  3.84\times 10^8\Ms$ is the total mass of the simulated galaxy region, and $M_{g,sim}/\langle m\rangle$ is the actual number of stars expected to orbit in the galaxy nucleus. We recall here that, due to the overwhelming computational load, only a fraction of the entire host galaxy has been modelled, corresponding to the nuclear region ($r\lesssim 150$ pc).

\begin{figure}
\centering
\includegraphics[width=8cm]{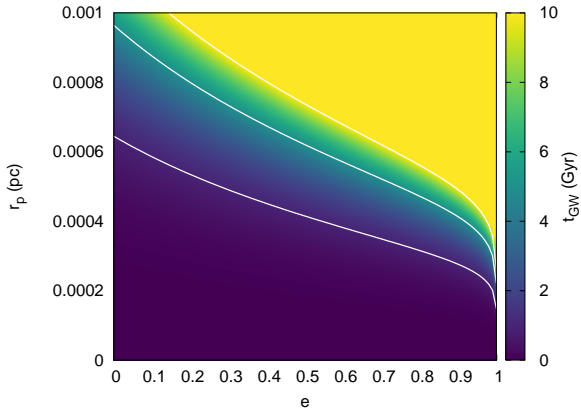}\\
\caption{GW time-scale as a function of the pericentre and eccentricity for an EMRI with total mass $10^8\Ms$ and mass ratio $10^{-7}$. The white straight line indicates a GW timescale of 1 (bottom line) 5 (central line) and 10 (top line) Gyr.}
\label{emrismap}
\end{figure}

Only a fraction of the newly formed EMRIs will merge within a Hubble time. 
The merger timescale of a binary system is well described by the \cite{peters64} formula
\begin{equation}
t_\gw =  \frac{5}{256}\frac{c^5 a_\bhb^4 (1-e_\bhb^2)^{7/2}}{G^3m_1m_2(m_1+m_2)},
\label{peters}
\end{equation}
where $c$ is the speed of light, $a_\bhb$ is the binary semi-major axis, $e_\bhb$ is the eccentricity, and $G$ is the gravitational constant.
Figure \ref{emrismap} shows how $t_\gw$ varies as a function of the EMRIs pericentre and eccentricity.

For instance an object having a pericentre $r_p \simeq 8\times 10^{-4}$ pc and eccentricity $e_\bhb \sim 0.4$ would undergo a merger within a Hubble time. Clearly, only sufficiently compact remnants will merge with the SMBH without undergoing tidal disruption, likely white dwarf (WDs), neutron stars (NSs) and BHs.  

Taking advantage of the SSE package \citep{hurley00}, conveniently modified to include the \cite{belckzinski10} prescription for stellar winds in heavy stars, we found that a fraction $f_{\rm bh} = 0.0024$ of the stars in a simple stellar population evolves in stellar BHs, while a fraction $f_{\rm ns} = 0.0094$ should evolve in NSs and $f_{\rm wd}= 0.12$ in WDs. Assuming a retaining fraction of $f_{\rm ret}\sim 100\%$ for all these populations and plugging the values in the equation above, we find a { number of EMRIs candidate }
\begin{equation}
N_{\rm EMRI} \simeq ( 4|_{\rm BH} + 15|_{\rm NS} + 195|_{\rm WD} +  ) = 214. \nonumber
\end{equation} 
In the calculation above we do not take into account mass segregation, which however crucially affects the mass distribution in the SMBH vicinity \citep{alexander09, seoane11, antonini14}, and can constitute the most common type of remnants in the stellar cusp around the SMBH.
In the most optimistic assumption that all the objects within $r_{\rm EMRI}$ are stellar BHs, we obtain instead
\begin{equation}
N_{\rm EMRI} \simeq 1659.
\end{equation}

The resulting EMRIs merger rate in the two limiting assumptions of no-mass segregation and full-mass segregation is thus given by:
\begin{equation}
\mathcal{R}_{\rm EMRI} = \frac{N_{\rm EMRI}(8\times 10^{-4})}{t_{\rm rel}} \simeq (0.2-1.7) \times 10^{-7} {\rm yr}^{-1}.
\label{emriE}
\end{equation}
Note that $\mathcal{R}_{\rm EMRI}$ does not depend significantly on the choice of the pericentral distribution, $f_1$ or $f_2$, because above $r_p=10^{-4}$ pc these two functions are similar. 

Our results suggest that if tidal forces are sufficiently strong to prevent the formation of a massive and dense nuclear cluster, the resulting low-density environment suppress EMRIs formation due to the smaller probability for close encounters to occur between compact objects and the SMBH.

Using the same procedure as above, we calculate the EMRIs rate at the beginning of the simulations, thus before the GCs infall and disruption affected the nucleus dynamics. Comparing $\mathcal{R}_{\rm EMRI}$ at different times, we found that the GCs evolution leads the rate to slightly increase by a factor $B \sim 1.2$.
It is worth noting that the deposit of GC debris produces an effect similar to the ordinary mass segregation process that occurs in a dense environment, but in this case it acts on a larger scale. 

{ In our models, the GC debris that can reach the galaxy innermost regions is comprised mostly of compact remnants (stellar BHs, white dwarves and neutron stars). Hence, the GCs infall drives the transport of a considerably heavy population of objects around the SMBH. In this sense, we can consider the GCs as a heavier component compared to the galaxy stars.
The GC debris slowly accumulates around the SMBH, leading to the formation of a steeper cusp compared to the ``unperturbed'' galaxy. Therefore, the combined action of dynamical friction and tidal disruption drives the accumulation of a noticeable amount of mass around the SMBH in form of stellar remnants, similarly to what is caused by mass segregation.}
\cite{seoane11} suggested that strong mass segregation can increase the EMRIs rate by a factor up to $10^4$. They parametrized the EMRIs rate through the adimensional parameter $\Delta$ defined as
\begin{equation}
\Delta = \frac{M_H \langle m_H \rangle}{M_L \langle m_L \rangle}\frac{4}{1+\langle m_H \rangle / \langle m_L \rangle},
\end{equation}
being $M_H$ the mass of the heavier component and $\langle m_H \rangle$ its average mass. The subscript $L$ indicates the lighter stellar component.
In our simulation, the GC debris mass enclosed within 10 pc is $\sim 85\%$ of the total mass orbiting the SMBH. { Since these debris are likely comprised of compact remnants, we can use the strong mass segregation approximation to calculate the enhancement in the EMRIs rate due to delivered objects. }

If we assume that the average stellar masses in the galaxy and the disrupted GCs are similar, $\langle m_H \rangle / \langle m_L \rangle\simeq 1$, we get $\Delta = 10$, to which corresponds an EMRIs rate boosting factor of $B \sim 1.5$, similar to the value calculated above.

The procedure described in this section relies on the assumption of a zero-spin SMBH. The overall picture is much more complex in the case of Kerr SMBHs. Indeed, the SMBH spin can affect significantly the location of the last stable orbit, allowing in some case orbiting stars to get closer to the SMBH and possibly enhancing the probability to get captured in an EMRI. 

As discussed by \cite{amaro13b}, the ratio $\mathcal{T}$ between the EMRIs rate for Kerr and Schwarzschild SMBHs is a function of the ratio between the last stable orbits $\mathcal{W}(s,i)$, which is a function of the SMBH spin ($s$) and the orbital inclination ($i$) of the star.
\begin{equation}
\mathcal{T} = \mathcal{W}(s,i)^{(20\gamma-45)/(12-4\gamma)},
\end{equation}
where $\gamma$ is the density slope around the SMBH. 

Assuming a nearly extremal SMBH ($s>0.999$) and moderate inclinations ($i\sim 20^\circ$), \cite{amaro13b} results indicate that $\mathcal{T} \simeq 5$ in the case of a steep density cusp ($\gamma=2$), while it can rise up to $\mathcal{T} \simeq 100$ if the density profile around the SMBH is smoother ($\gamma \simeq 0.5$), like in our simulations.

As a consequence, low-density galactic nuclei harbouring a rapidly rotating SMBH can be characterized by an EMRIs rate up to 100 times larger than for a non-rotating SMBH, { although the actual boosting factor depends on the parameters distribution. This is likely connected to the distribution function that governs the dynamics (Pau Amaro-Seoane, private communication)}.

The results discussed above can be used to infer the rate of EMRIs formation and coalescence in the local Universe. A crucial parameter to perform such a calculation is the number density of galaxies similar to the MEGaN model, $n_g$, at redshift $z=0$. We recall here that our simulation is tailored to a galaxy with total mass of $M_g = 10^{11}\Ms$, corresponding to a stellar mass $M_{g*} \simeq 6.3\times 10^{10} \Ms$ \citep{Gallazzi2006}. Hence, taking advantage of the Illustris\footnote{\url{http://www.illustris-project.org/}} simulation data release \citep{Nelson15}, which provided a simulation of structures formation at an unprecedented level of detail, we calculated for galaxies with a stellar mass above $M_{g*} \simeq 6.3\times 10^{10} \Ms$, a galaxy number density $n_g \sim 10^{-3}$ Mpc$^{-3}$.

Combining these quantities with Eq. \ref{emriE}, we obtain a total EMRIs rate for massive ellipticals in the local Universe of
\begin{equation}
\Gamma_{\rm EMRI} = \mathcal{R}_{\rm EMRI} n_g = (0.02 - 0.17) ~{\rm yr}^{-1} ~{\rm Gpc}^{-3}.
\end{equation}

\section{From ARCHAIN to ARGdf: combining algorithmic regolarization with dynamical friction}
\label{ARGdf}

In this section we will investigate the possibility that the infalling clusters deliver to the galactic centre either an IMBH or a population of stellar BHBs. 

Understanding the efficiency of IMBH-SMBH pairing in consequence of GC disruption is of paramount importance in the perspective of the next generations of space based GW detectors, such as LISA, TianQin and Taiji \citep{gualandris09,baumgardt06,mastrobuono14,fragione17c,ASG17}.

On another hand, quantifying the role of BHB-SMBH encounters in the development of GWs by stellar BHBs can allow to better constrain the origin of the known population of merged BHBs as seen by the LIGO/VIRGO collaboration \citep{abbott16a,abbott16b,abbott16c,abbott17a,abbott17d}.

The current version of the \HiGPUs code used to perform the MEGaN simulation, allowed us to model a total number of particles sufficiently large, $N=2^{20}\sim 10^6$, to ensure a correct representation of both the ``internal'' and orbital evolution of the GCs.
Recent updates to the code allowed us to include a special control for SMBH binaries \citep{ASGER18} and to take into account single stellar evolution \citep{AS16}, thus representing potential improvements for future numerical models. 

Although the MEGaN simulation represents a state-of-art in the field of direct $N$-body modelling of galaxy nuclei, its resolution was still far from a one-to-one representation. Indeed, each particle in our GC models has a mass $\sim 100\Ms$, larger than the heaviest known stars.
None of the currently available \HiGPUs version treat propertly the dynamics of tight systems, such as binaries and triples, making impossible to model the complex subsystems that are expected to form during the GC evolution. Moreover, \HiGPUs does not include general relativity effects, which are crucial to follow the last phases of BHs evolution in binary systems.

Unfortunately, a direct summation code that allows modelling a whole galaxy nucleus and, at the same time, can integrate carefully the pc-scale evolution of IMBHs and BHBs, does not exist. A few examples have been provided recently \citep{khan16,haster16,hayasaki18}, although none of them take into account all the features at the same time .

Due to this, we combined the detailed information provided by the MEGaN model with our code \ARGdf.
As discussed in the following, \ARGdf allows us to obtain a detailed representation of the evolution of IMBHs and BHBs around an SMBH.

Based on the Mikkola's \texttt{ARCHAIN} code, \texttt{ARGdf} is a few-body integrator which implements the algorithmic regularization scheme, a treatment adapted to model strong gravitational encounters \citep{mikkola99} including post-Newtonian terms up to the 2.5th order \citep{mikkola08}. Our new implementation, called \texttt{ARGdf}, takes into account both the gravitational field of the galaxy, treated either as a \cite{Deh93} or a \cite{Plum} analytic field, and the dynamical friction (df) effect.

Using \ARGdf, in the following we will explore the possibility that some of the infalling clusters transport to the galactic centre either an IMBH or several stellar mass BHBs.

In the first scenario, a reliable representation of the IMBH orbit requires a correct evaluation of the dynamical friction acting on the IMBH. On another hand, to properly reproduce BHBs around an SMBH the code must cope the difficulty to model the BHB internal evolution, on AU-scale, and the BHB revolution around the SMBH, on pc-scale, at the same time.

\subsection{Dynamical friction term for self-interacting massive black holes}

As pioneered by \cite{Cha43I}, the df term acting on a point-like satellite with mass $M$ orbiting at distance $r_M$ from the centre of the hosting system and having speed $v_M$ is given by
\begin{equation}
\label{chandradf}
\frac{{\rm d}{ v}}{{\rm d}t} \simeq -4\pi^2G^2 M\ln \Lambda \rho(r_M,v_m\leq v_M)\frac{{ v}_M}{v_M^3}.
\end{equation}
Note that here, $\rho(r_M,v_m\leq v_M)$ is the local density of field stars slower than the satellite. 
In the simplest approximation that the galactic distribution function can be described by a Maxwellian, the df term can be calculated as \citep{bt}
\begin{equation}
\frac{{\rm d}{ v}_{M}}{{\rm d}t} = -4\pi G^2 M \rho \ln\Lambda \mathcal{F}\frac{{ v}_M}{v_M^3}.
\label{chandra}
\end{equation}
with 
\begin{equation}
\mathcal{F}\equiv \mathcal{F}\left(\frac{v_M}{\sqrt{2}\sigma}\right)= \left[\erf\left(\frac{v_M}{\sqrt(2)\sigma}\right) - \frac{2v_M}{\sqrt{2\pi}\sigma}\exp(-v_M^2/2\sigma^2)\right].
\end{equation}

A severe limitation in the use of this formula arises when the satellite moves at low velocities in the inner region of the host system, situation that implies $|{\rm d}{ v}_M/{\rm d}t|\rightarrow \infty$. 

One of the first refinement to Chandrasekhar's work was proposed by \cite{binney77}, treating dynamical friction in aspherical systems \citep[see also][]{penarrubia04}.
Later on, much work has been done to provide a reliable treatment of dynamical friction in cuspy systems hosting central SMBHs \citep{Just11,AntMer12,ASCD14a,Petts15,Petts16} or in triaxial galaxies \citep{OBS,Pes92,Dosopoulou17}.

Moreover, as discussed by \cite{ASCD14a}, the interactions between two massive satellites in the galactic centre can result in a further orbital energy loss that yields to a faster orbital decay (see also \cite{Dosopoulou17}).

In the MEGaN simulation, the dynamical friction acting on the infalling GCs is a natural outcome of dynamics, and the corresponding spiralling trajectory does not suffer of any approximation but the coarse grained nature of the simulation, which slightly differ from a real galaxy since our model is comprised of a much smaller particles number.

Hence, to provide a correct modelling of the evolution of one, or more, IMBHs delivered from infalling GCs around an SMBH, we need to refine Eq. \ref{chandra}. 

To characterize this problem, we ran three different simulations representing multiple MBHs orbital decay in a dense environment. 

We used \HiGPUs to create three self-consistent models of a galactic nucleus using $131072$ particles, where we injected up to 3 MBHs, varying for each simulation their initial conditions. In order to reduce the computational time, we decided to set in this case $M_g = 10^{11}\Ms$ for the galaxy mass, and $r_g = 200$ pc. This choice leads to shorter df time-scales, thus allowing to gather a few test-runs at a reasonable computational cost. In these models, we set the softening parameter to $10^{-4}r_g$. The df acting on the MBHs will come out naturally from their interaction with field stars, thus representing the ideal reference frame to calibrate \ARGdf. 

In the following, we refer to these test simulations as T1, T2 and T3. Their main properties are summarized in table \ref{ARGDFsim}.

\begin{table*}
\caption{}
\centering{Main parameters of \ARGdf test runs}
\begin{center}
\begin{tabular}{cccccccccc}
\hline
ID & $M_1$  & $r_1$ & $e_1$ & $M_2$ & $r_2$ & $e_2$ & $M_3$ & $r_3$ & $e_3$\\
   & $(M_g)$& $r_g$ &     &$(M_g)$& $r_g$ &     &$(M_g) $& $r_g$ &    \\
\hline
$T1$ & $10^{-3}$& $5\times 10^{-2}$ & $0.8$ & $10^{-3}$& $5\times 10^{-2}$ & $0.8$ & - & - & - \\
$T2$ & $10^{-3}$& $10^{-2}$ & $0$ & $10^{-5}$ & $1$ &$ 0.68$ & - & - & - \\
$T3$ & $10^{-3}$& $5\times 10^{-2}$ & $0$ & $10^{-3}$& $5\times 10^{-2}$ & $0$ & $10^{-3}$ & $0$ & - \\ 
\hline
\end{tabular}
\label{ARGDFsim}
\end{center}
\end{table*}

In simulation T1, we modelled the evolution of two equal mass SMBHs with masses $M_1 = M_2 = 10^{-3} M_g$, initially placed symmetrical with respect to the galaxy centre and moving on an eccentric orbit with eccentricity $e_1 = e_2 \simeq 0.8$ and initial apocentre $r_{1,2} = r_g/20$.  

In simulation T2, instead, we put a lighter MBH, with mass $M_2=10^{-5}M_g$, on an eccentric orbit ($e_2=0.68$) at a distance $r_2 = r_g$ from the galactic centre, while a heavier MBH, $M_1 = 0.001M_g$, moves on a tighter circular orbit, $r_1 = 0.01r_g$. With these assumptions we can investigate how the inner and heavier SMBH perturbs the evolution of the outer, lighter, MBH.

Finally, in simulation T3 we modelled the evolution of three MBHs with identical masses, $M_1 = M_2 = M_3 = M_g/1000$, two of them initially placed symmetrically with respect to the galactic centre at $r_{1,2} = r_g/20$, while the third is set in its centre.
The two BHs, with initial separation $r_{1,2} = 2r_{1}$, move on initially circular orbits.

Note that in model T3 we started our \ARGdf~ simulation by using the \HiGPUs~taken at $\sim 0.1$ Myr as input conditions, in order to take into account the initial displacement of the central MBH from the galactic centre, unavoidable in the full N-body models, due to its interaction with the field stars.

\begin{figure}
\centering
\includegraphics[width=8cm]{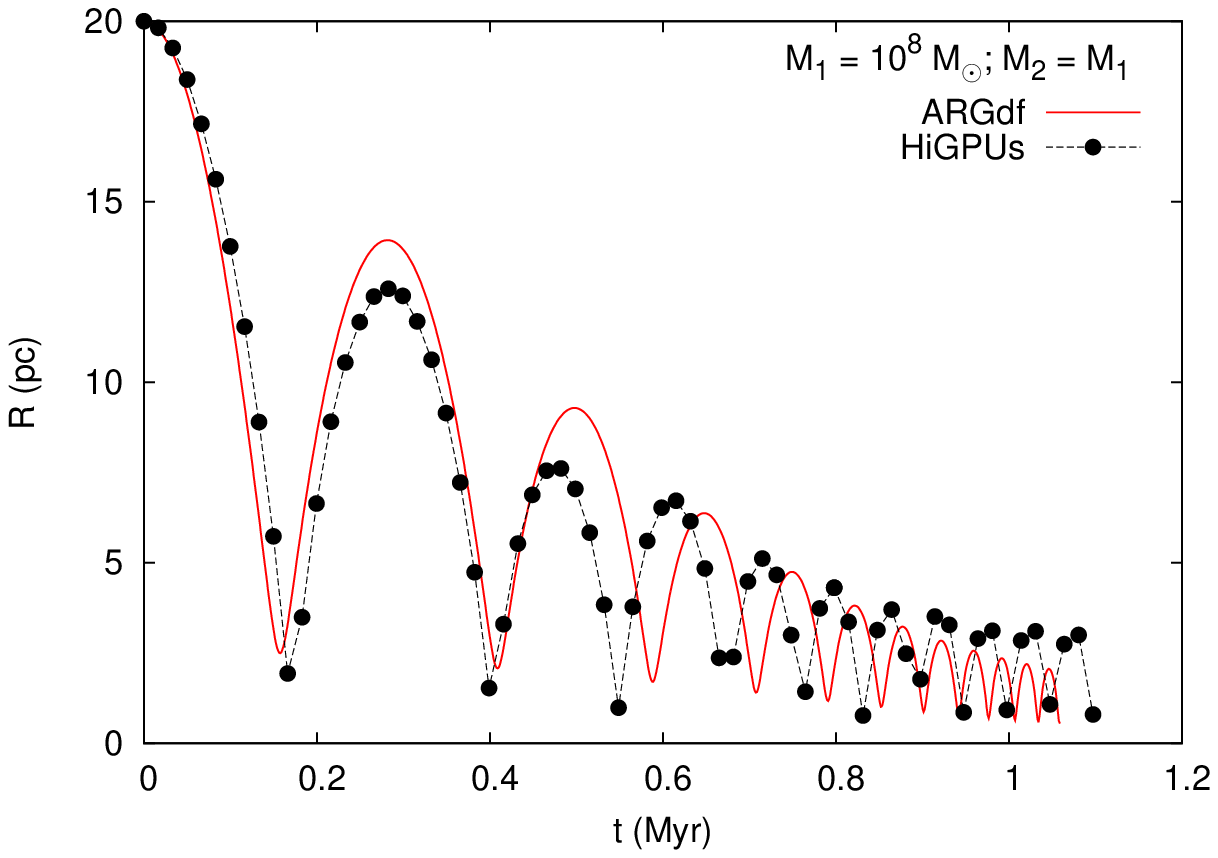}\\
\includegraphics[width=8cm]{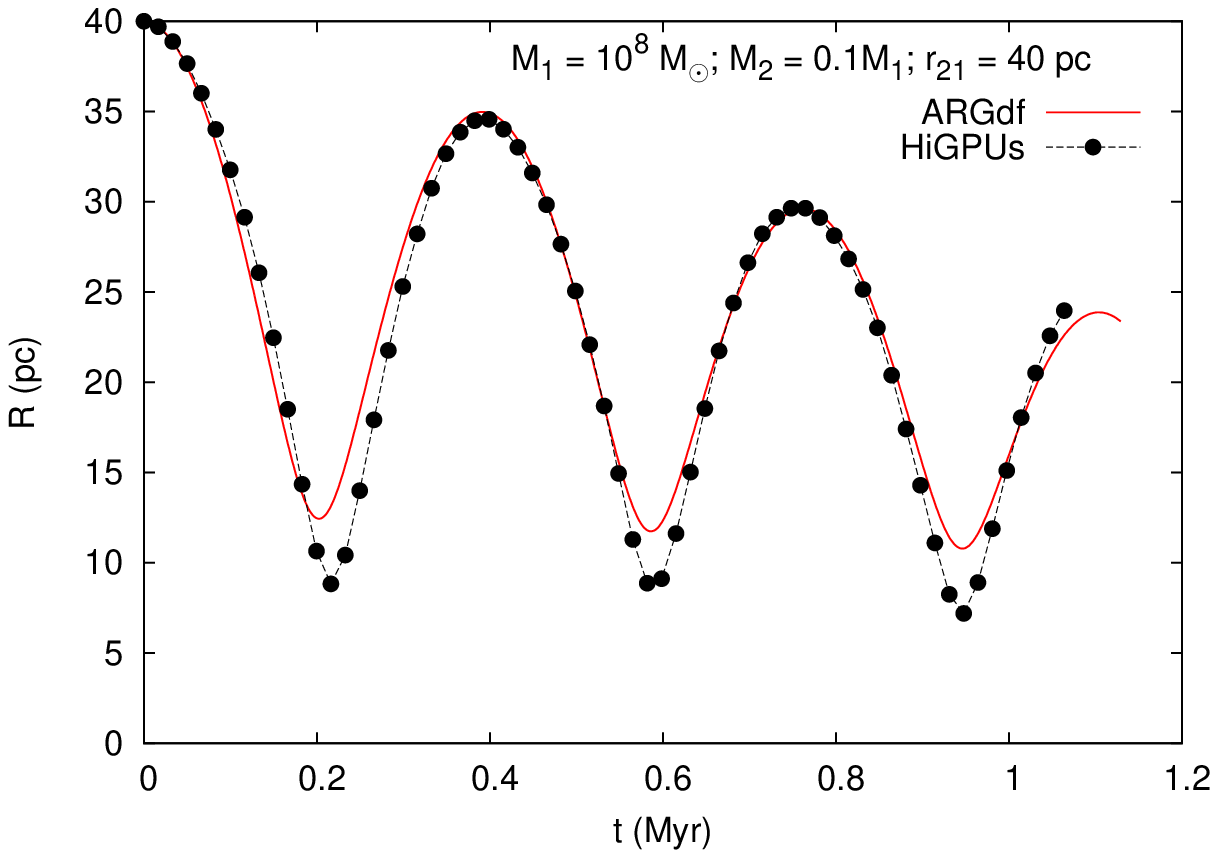}\\
\includegraphics[width=8cm]{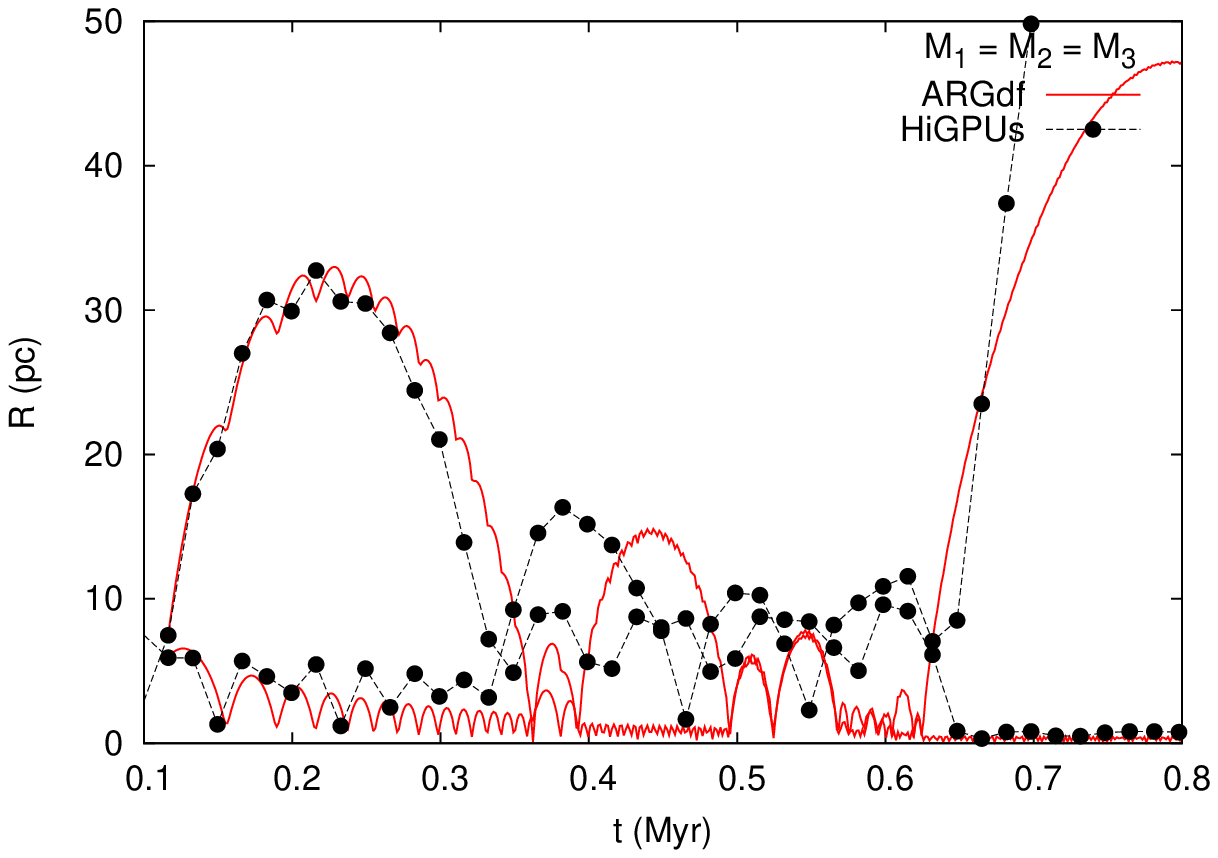}
\caption{Top panel: mutual distance between SMBHs $M_1$ and $M_2$ in model T1 while travelling in a ``live'' galaxy (\HiGPUs, black dotted line) and in an external potential with the additional df term (\ARGdf, red straight lines).  Central panel: radial distance to the galactic centre for SMBH $M_2$ in model T2. Bottom panel: time evolution of the radial distance to $M_3$ for $M_1$ and $M_2$ in model T3. }
\label{ARGdfcmp}
\end{figure}

Using the HIGPUs runs as a comparison test-bed, we sought for an optimal generalization of the local approximation formula (Eq. \ref{chandradf}) in the form of a multiplicative \textit{shape function} $\mathcal{L}$. As discussed in the next Section \ref{appA}, it turned out that $\mathcal{L}$ depends on the MBH masses and on the galaxy background properties. 

Figure \ref{ARGdfcmp} shows how the MBHs mutual distances vary in all the simulations performed with our few-body approach and the full N-body code. Although the \HiGPUs and \ARGdf results do not completely overlap, our approach seems to provide a reliable approximation to the overall evolution of multiple MBHs in a dense environment.

\subsection{A shape function for the dynamical friction suffered by multiple MBHs}
\label{appA}

As indicated in Section \ref{ARGdf}, to represent correctly the effect of dynamical friction in our few-body code \ARGdf we need to modify the standard \cite{Cha43I} formula (Eq. \ref{chandradf}). This is done by multiplication of the dynamical friction term in Eq. \ref{chandradf} by a \textit{shape factor}, which is a function accounting for the perturbations induced by the other massive satellites, and that avoids the divergence for the satellite velocity approaching zero.

After a careful comparison between the full N-body simulations, comprised of $N=131k$ particles and performed with HIGPUs, we found that the shape function $\mathcal{L}$ can be defined as:
\begin{equation}
\mathcal{L} = \left[1+\left(\frac{\rho_{\rm bh,i}}{\rho_g(r)}\right)\right]D(M_{\rm tot},\mu_{\rm tot}) .
\label{shape}
\end{equation}

Here, $\rho_g(r)$ is the galaxy density, calculated at the position of the infalling satellite, while $\rho_{\rm bh,i}$ is a corrective factor given by
\begin{eqnarray}
\rho_{\rm bh,i} =& M_i/R_{i}^3,\\
R_i         =& r_g \frac{\left(M_i/M_g\right)^{1/(3-\gamma)}}{1-\left(M_i/M_g\right)^{1/(3-\gamma)}},
\end{eqnarray}
note that $R_i$ represents the so-called stalling radius, 
i.e. the distance to the galactic centre where dynamical friction efficiency ceases \citep{readcole,AntMer12}, which roughly coincides with the radius at which the galaxy and the satellite masses equal each other \citep{gualandris08,ASCD14a}. 

As discussed in \cite{ASCD14a}, the interaction between two massive satellites can significantly affect their orbital decay. They suggested that the satellites can undergo a ``super-segregation'' phase as long as their mass ratio is below $\lesssim 10-100$. After the first close interaction, the satellite apocentre can decrease up to 3 times the value achieved in absence of massive perturbers, leading the dynamical friction time-scale to decrease by roughly the same factor. The energy transfer between the satellites and the background stars will be efficient as long as the satellite approaches the stalling radius, where the satellite self-interaction start dominating over dynamical friction. In these regards, the $\rho_{\rm bh,i}$ term mimics the effect above, enhancing the dynamical friction term and allowing to correctly reproduce the apocentre decrease observed in the full $N$-body models.

This damping mechanism can occur each time that two satellites undergo a close interaction. Hence, multiple interactions can further enhance the orbital segregation. The number of possible interaction for each satellite with the others is $N-1$, while the amplitude of the resulting damping will depend on the satellites mass ratio. In order to account for this effect, we introduced the second term in equation \ref{shape}, defined as
\begin{equation}
D(M_{\rm tot},\mu_{\rm tot}) = \frac{1}{20} (N-1) \left(\frac{M_{\rm tot}}{\mu_{\rm tot}}\right)^{1/3},
\end{equation}
where
\begin{eqnarray}
M_{\rm tot}   &=& \sum_{i=1}^n M_i, \\
\mu_{\rm tot} &=& \displaystyle{\frac{\Pi_{i=1}^n M_i}{\left(\sum_{i=1}^n M_i\right)^{n-1}}},
\end{eqnarray}
represent the satellites' total and reduced masses, respectively. This term is constant for all the satellites in the sample, and allow to obtain an average friction enhancing factor that mimics the effects of satellites self-interactions. 

As widely discussed in literature, \cite{Cha43I} formula overestimate the actual frictional term as long as the satellite velocity approaches zero \citep{Just05,vicari07,Just11,ASCD14a,Petts15,Petts16}.
Following \cite{Webb18}, we replaced the satellite velocity $v_M$ in equation \ref{chandra} with $\left(v_M + \sigma_{\rm tot}\right)$, where the additional term is given by
\begin{equation}
\sigma_{\rm tot}^2 = \frac{GM_g(r)}{r} + \frac{GM_i}{R_i}.
\end{equation}
Note that for a \cite{Deh93} model, the first term goes to zero as $r^{2-\gamma}$, while the second term is a constant that accounts for the fact that dynamical friction inside the stalling radius drops to zero.

\section{Results from few-body models: IMBHs and BHBs merging in galactic nuclei}
\label{results}

In this section, we will use \ARGdf to study the interactions between the MEGaN central SMBH and several IMBHs and BHBs possibly delivered by a population of orbitally segregated GCs.

Regarding IMBHs, we explore several scenarios: i) 2 IMBHs, initially hosted in the two most massive GCs in our sample, start interacting with the SMBH after the GCs disruption; ii) 1 IMBH orbited by a BHB heading to the SMBH; iii) 8 IMBHs are left around the SMBH by their progenitor GCs.

Regarding BHBs, instead, we will investigate whether the SMBH gravitational pull can facilitate the shrinking process and drive the binary to the coalescence.

We calculate the expected rate at redshift $z=0$ for all the channels above. 

In order to quantify the events rate $\Gamma$, we need several ingredients: i) the probability for such an event to occur, $P_{\rm mer}$; ii) the number of GCs that are expected to segregate into the galaxy nucleus $n_{\rm dec}$; iii) the number of BHBs expected to reside in the GC during the galaxy nucleus assembly $n_{\rm BHB}$; iv) the number density of massive galaxies in the local Universe $n_g$; v) the time-scale over which the merger event occurs $t_{\rm mer}$.

Once all these ingredients are defined, the rate at which a merger occurs in one of the channel listed earlier can be described by a Drake-like function
\begin{equation}
\Gamma = P_{\rm mer}n_{\rm dec}n_{\rm BHB}n_gt_{\rm mer}^{-1}.
\label{EQgamma}
\end{equation}

Table \ref{rates} summarizes our results at this regard.

\begin{table}
\caption{Merger rates for the different channels investigated}
\begin{center}
\begin{tabular}{ccc}
\hline
objects pair   & location & $\Gamma$ \\
               &          & yr$^{-1}$ Gpc$^{-3}$\\
\hline
BH-SMBH  & Galaxy nucleus & 1.33\\
IMBH-SMBH& Galaxy nucleus & 0.03\\
BH-BH    & around the IMBH& 2.0 \\
BH-IMBH  & GC nucleus     & 9.5 \\
BH-BH    & Galaxy nucleus & 1   \\
\hline
\end{tabular}
\end{center}
\label{rates}
\end{table}

\subsection{Multiple IMBHs-SMBH scattering: formation of IMRIs}
\label{IMBH}

The orbital evolution of several IMBHs left into the galactic nucleus after the disruption of their progenitor cluster can be divided into three different phases:
\begin{itemize}
\item the IMBH is carried by the GC, whose orbital pericentre reduces substantially over a df time-scale $\tau_{\rm df}\propto M^{-0.67}r_{p0}^{1.74}$. Note that here $M$ is the GC mass;
\item the GC is disrupted by tidal forces after a few passages at pericentre, leaving the IMBH freely wandering within the galactic nucleus;
\item the dynamical friction exerted on the IMBH by the field stars causes its orbital decay toward the central SMBH, on a time-scale $\tau_{\rm df} \propto {M_\ibh}^{-0.67}{R_\ibh}^{1.74}$, being $R_\ibh$ the IMBH distance to the SMBH.
\end{itemize}

Clearly, the number of IMBHs possibly inhabiting a galactic center depends critically on the IMBH formation success rate. 
For instance, the absence of IMBHs observational signatures around the MW SMBH implies either that the formation of an IMBH in a GC is an unlikely event, that they have already merged with the central SMBH, or that the time-scale for repeated GCs mergers is much shorter than the IMBH formation time.

\citet{Giersz15} computed 2000 Monte Carlo simulations of GCs, showing that the formation of an IMBH is an extremely {\it stochastic} event, strongly dependent on the GC initial mass and properties. In their sample, an IMBH forms in $\sim 20\%$ of the cases. In our sample, nearly 10 clusters out of 42 have eccentricities above $0.6$ and a mass above $10^6\Ms$, which, according to \citet{Giersz15}, would translate into only 2 IMBHs expected. Therefore, we followed the evolution of 2 IMBHs dragged toward the SMBH by GC infall in order to understand whether their following evolution can lead to an SMBH-IMBH merger within a Hubble time.

The IMBH masses were assigned according to the scaling law proposed by \cite{AS16}, which links the mass of the central compact object sitting in the GC centre (either an IMBH or a subsystem of mass-segregated stellar BHs \cite[see also][]{AAG18a} to the total GC mass, that is

\begin{equation}
{\rm Log} M_\ibh = a {\rm Log} M + b,
\label{scale}
\end{equation}

where the coefficients $a$ and $b$ depend on the IMF and the metallicity of the host GC.

For a low-metal cluster characterised by a Kroupa IMF, \cite{AS16} found $a = 0.999$ and $b = -2.238$. This estimate is in good agreement with the general results found in \cite{zwart07} and with observational correlations provided by \cite{Lutzgendorf13} on the basis of several observed putative IMBHs.

The heaviest GCs in our sample have masses $M_{\rm GC1} = 2\times 10^6 \Ms$ and $M_{\rm GC2} = 1.9\times 10^6 \Ms$ respectively. Hence, according to Eq. \ref{scale}, the corresponding IMBH mass is $M_\ibh = 1.12\times 10^4\Ms$ and $1.08\times 10^4\Ms$, respectively.

In the MEGaN model, the heaviest GC, $M_{\rm GC1}$, moves on an orbit closer to the SMBH, thus approaching the galaxy centre earlier. Due to this, we assumed that its IMBH segregates faster than the other and binds to the SMBH, forming a binary system moving in a circular orbit with separation $\sim 1$ pc. 
This assumption is justified by that, in a nearly spherical configuration, the inner binary initially shrinks due to the interaction with field stars, circularizes and eventually stalls due to the inefficient replenishment of the so called loss-cone. 
The coalescence time-scale in such configuration, calculated through Eq. \ref{peters}, is enormous, $\sim 10^{35}$ yr.

We simulated the evolution of the triple system composed by the SMBH-IMBH pair and the outer IMBH taking into account the galactic tidal field and the df coefficient through the \ARGdf code, as discussed in Section \ref{ARGdf}.

However, after the second IMBH is left to freely move into the galactic nucleus, it is impossible to know what is the mutual inclination and orientation of its orbit with respect to the IMBH-SMBH pair sitting in the galactic centre.

The two IMBHs and the SMBH form rapidly a triple system, whose evolution can be easily followed by defining an inner and an outer binary. Initially, the inner binary is comprised of the SMBH and the heavier IMBH, whereas the outer binary is represented by the inner binary centre of mass and the third IMBH. 

In order to cover our ignorance about the outer IMBH orbital properties, we performed 100 simulations at varying the outer IMBH eccentricity, which is in part inherited from the parent GC orbit, and the inclination angle between the inner and outer binary orbits.

Therefore, in our simulations the inner binary has a total mass $M_{\rm in} = \sim 10^8\Ms$ and mass ratio $q_\bhb = 10^{-4}$, while for the outer binary, with mass $M_{\rm out} \sim 10^{-4} M_{\rm in}$, we varied semi-major axis ($a_{\rm out} = 2-40$ pc), eccentricity ($e_{\rm out} = 0.5-1.0$) and orbital inclination ($i=0-180^\circ$).

We found a $51\%$ probability for the inner binary to undergo coalescence within 14 Gyr. More in detail, in $\sim 38\%$ of the models the secondary IMBH substitutes the primary
and drives the formation of a tighter inner binary that undergo coalescence within a Hubble time. The semi-major axis shrinks down to AU-scales, leading the IMBH-SMBH binary in the GW dominated regime. The eccentricity undergo complex oscillations during the chaotic interactions that characterize the swapping phase. After the formation of the new, tight, binary, the eccentricity slowly decreases down to zero as a consequence of GW emission.
The evident sharp transition in the evolution of the semimajor axis in all the 3 panels marks the moment in which the component exchange takes place. 

In the remaining $14\%$ of the models, instead, the merger is boosted by the perturbations induced by the outer IMBH.

\begin{figure}
\centering
\includegraphics[width=8cm]{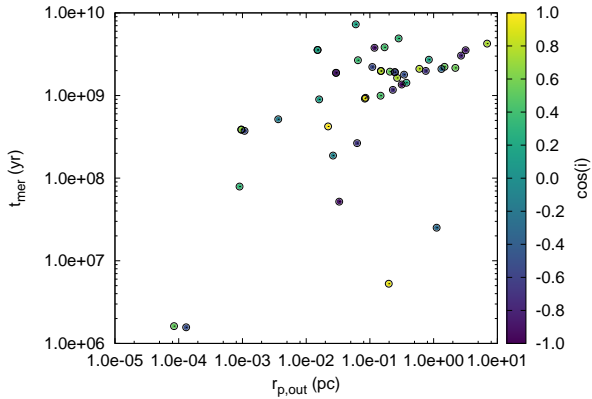}
\caption{Time at which an IMBH-SMBH merger occurs as a function of the outer binary initial pericentre. Different colours label the initial mutual inclination between the inner and the outer binary orbital planes. }
\label{bhbmer}
\end{figure}

Figure \ref{bhbmer} shows the time at which the merger occurs as a function of the outer binary initial pericentre. The plot also outlines the initial inclination between the inner and outer binary.

It seems evident a weak dependence of the coalescence time by the outer binary initial pericentre, while no correlation with its inclination is seen.

\begin{figure}
\centering
\includegraphics[width=8cm]{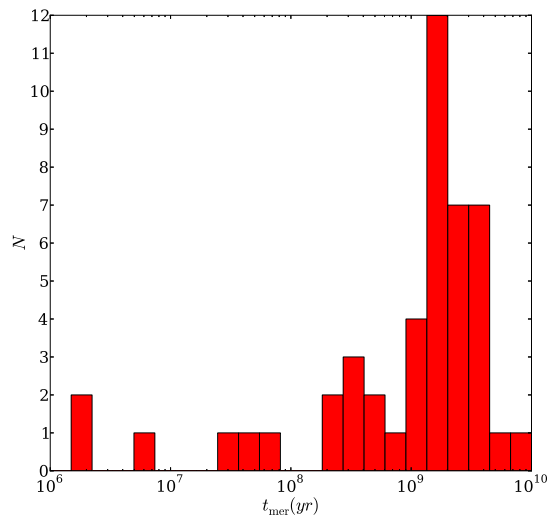}
\caption{Distribution of merger times in the 2IMBH+SMBH models set.}
\label{merhisto}
\end{figure}

Our results suggest that the probability for a triple system comprised of 2 IMBHs and a SMBH to be the site of a merger is quite large, $\sim 50\%$, quite independently on the IMBHs initial conditions. The time at which the merger occurs, $t_{\rm mer}$, clearly depends on the outer IMBH initial orbit. 
Figure \ref{merhisto} shows the $t_{\rm mer}$ distribution. In $73\%$ of the cases, the merger occurs on a time-scale longer than $1$ Gyr and is mediated by an outer IMBH with initial pericentre $r_{\rm p, out}>10^{-2}$ pc. In the remaining $27\%$ the merger is faster, with merging times ranging between $10^6-10^9$ yr. We note that nearly half of the coalescences develop after 2-3 Gyr since the beginning of our simulations.

In 35 models the outer IMBH unbinds from the inner binary after a few interactions, being ejected either into the galactic bulge or even away from the galaxy.
In 9 cases, the close interaction between the inner and outer binary leads both to the ejection of the outer IMBH and the merger of the inner SMBH-IMBH system.

\begin{figure}
\centering
\includegraphics[width=8cm]{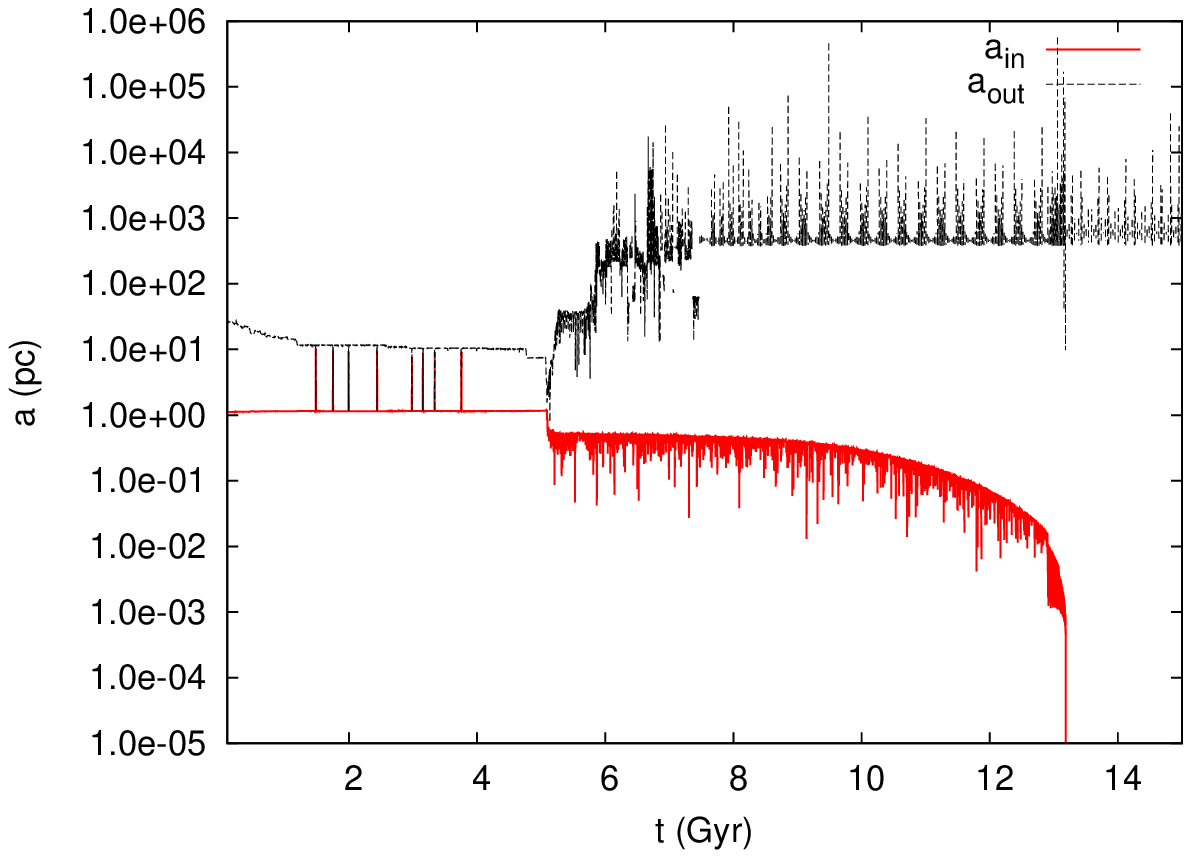}\\
\includegraphics[width=8cm]{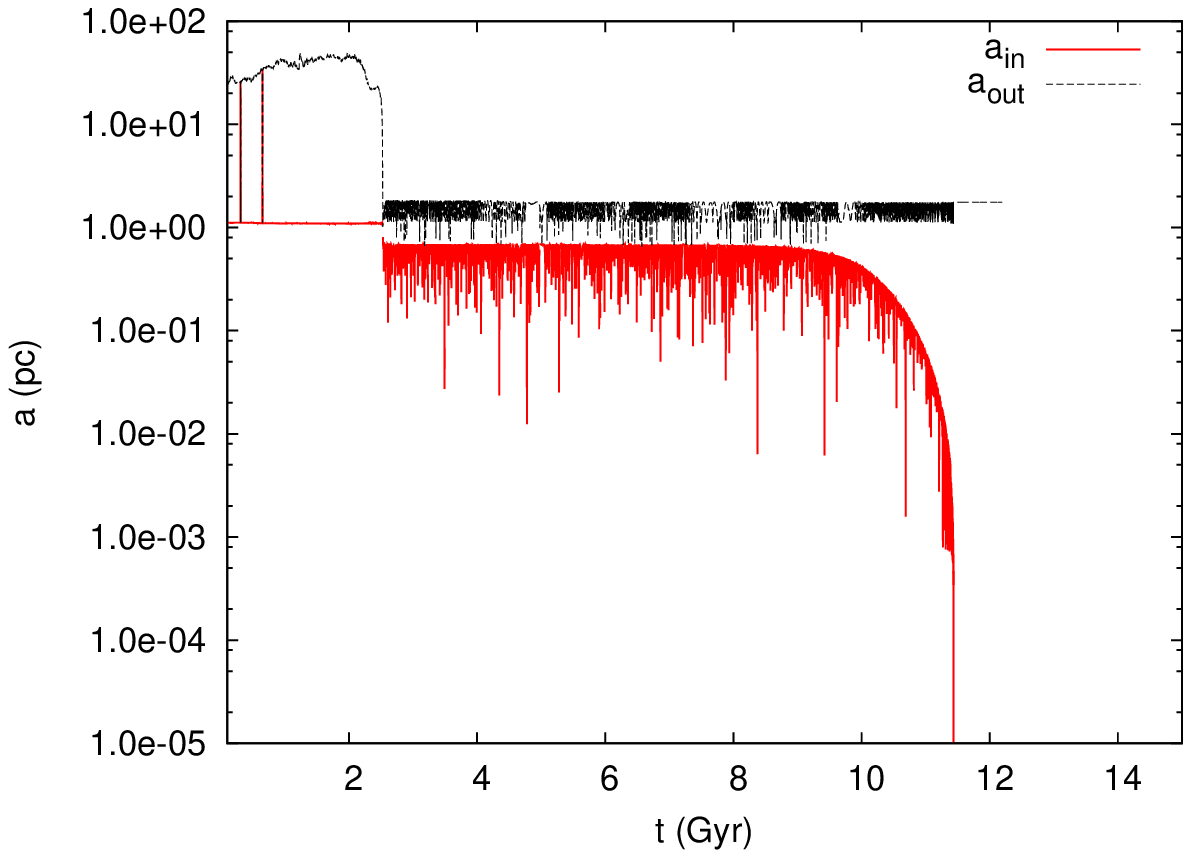}\\
\includegraphics[width=8cm]{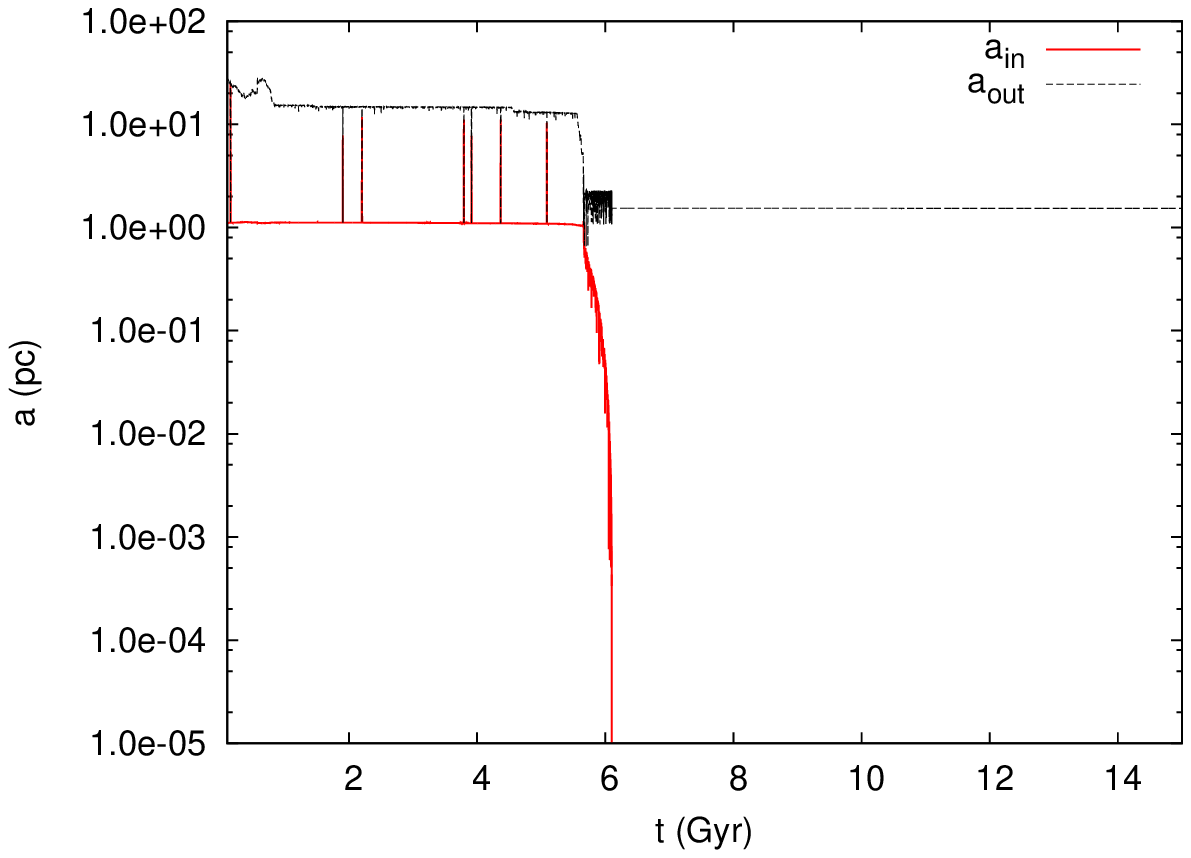}
\caption{The three panels represent the time evolution of the inner (red straight line) and outer (black dotted line) binary semi-major axis for three out of the 100 simulations of triple (SMBH + 2 IMBH) systems.}
\label{IMBHmot}
\end{figure}

\begin{figure}
\centering
\includegraphics[width=8cm]{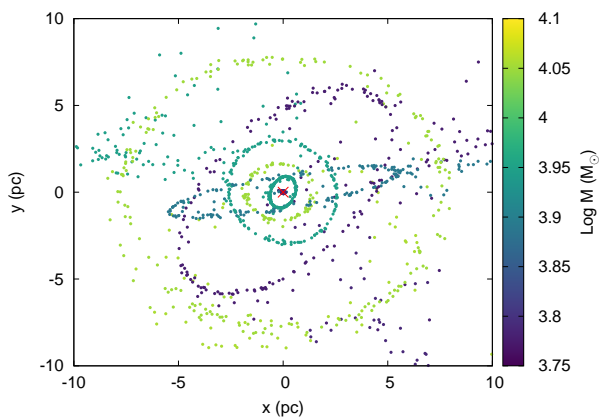}
\caption{Trajectories of 7 IMBH in one of the modelled systems. The central red cross represents the SMBH.}
\label{traj}
\end{figure}

For the sake of comparison, we also run several models in which the $20\%$ of the whole GC population hosts an IMBH in its centre, thus implying 7 IMBHs orbiting around the central SMBH. 
Figure \ref{traj} shows an example of IMBHs trajectories in such a case.

We kept the same inner binary as above, composed of an IMBH with mass $M=1.12\times 10^4\Ms$ moving on a circular orbit around the central SMBH, while the other IMBHs are distributed randomly in the space, since it is impossible to predict their exact position and velocity shortly after the host GC disruption. In order to focus on the IMBHs-SMBH interactions, we set 50 times the inner binary separation as the maximum distance from the central SMBH allowed.

We gathered 474 models, divided into 4 different groups, which differ each other in two important features: i) dynamical friction and ii) central SMBH spin.
Regarding the first point, a half of the models were run taking into account the background dynamical friction. Regarding the second point, instead, in a half of the models we assigned to the central SMBH an adimensional spin parameter $s = 0.1$, value corresponding to $10\%$ of the maximal spin allowed for a Kerr BH. Such a value is compatible with spins of SMBHs with mass above $4\times10^7\Ms$, as inferred by observations of AGNs \citep{reynolds13}.

We set the origin of the reference frame in the inner binary centre of mass, and the xy plane is assumed to coincide with the inner binary orbital motion. According to this configuration, we arbitrarily oriented the SMBH spin vector aligned to the y-axis.

In the following, we refer to the 4 groups of simulations as DFnSn, DFySn, DFnSy, DFySy. 
Here, the subscript ``y'' stands for ``yes'' and ``n'' for ``no'', thus allowing to easily find which process is taken into account in each model set.

\begin{figure}
\centering
\includegraphics[width=8cm]{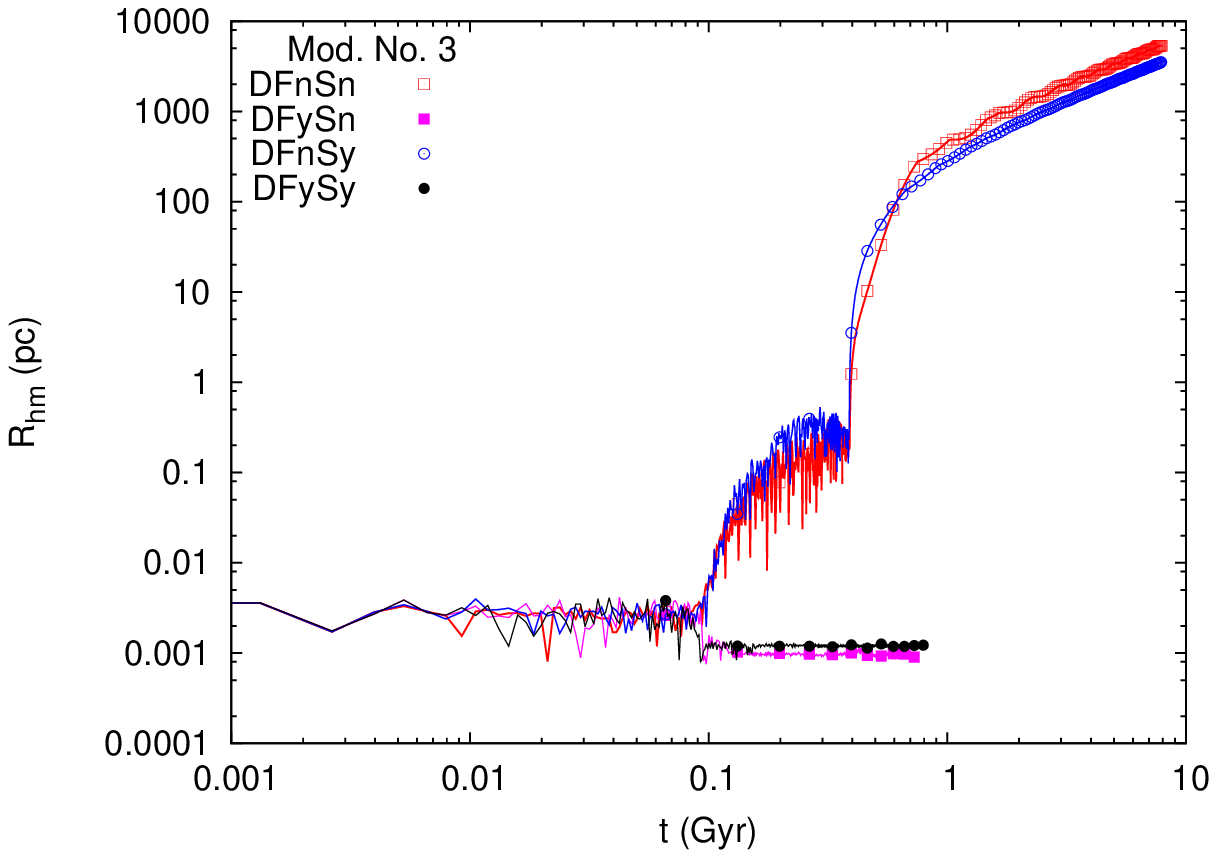}\\
\includegraphics[width=8cm]{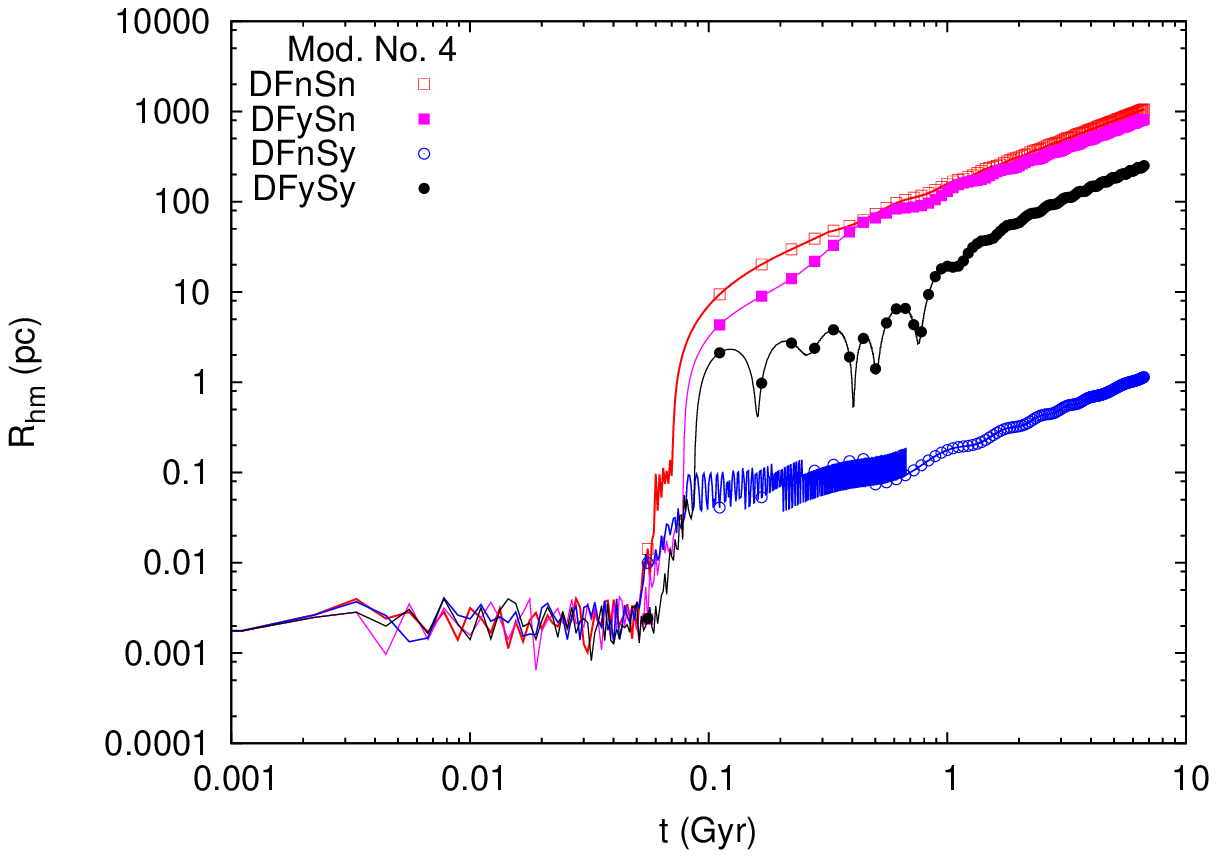}\\
\includegraphics[width=8cm]{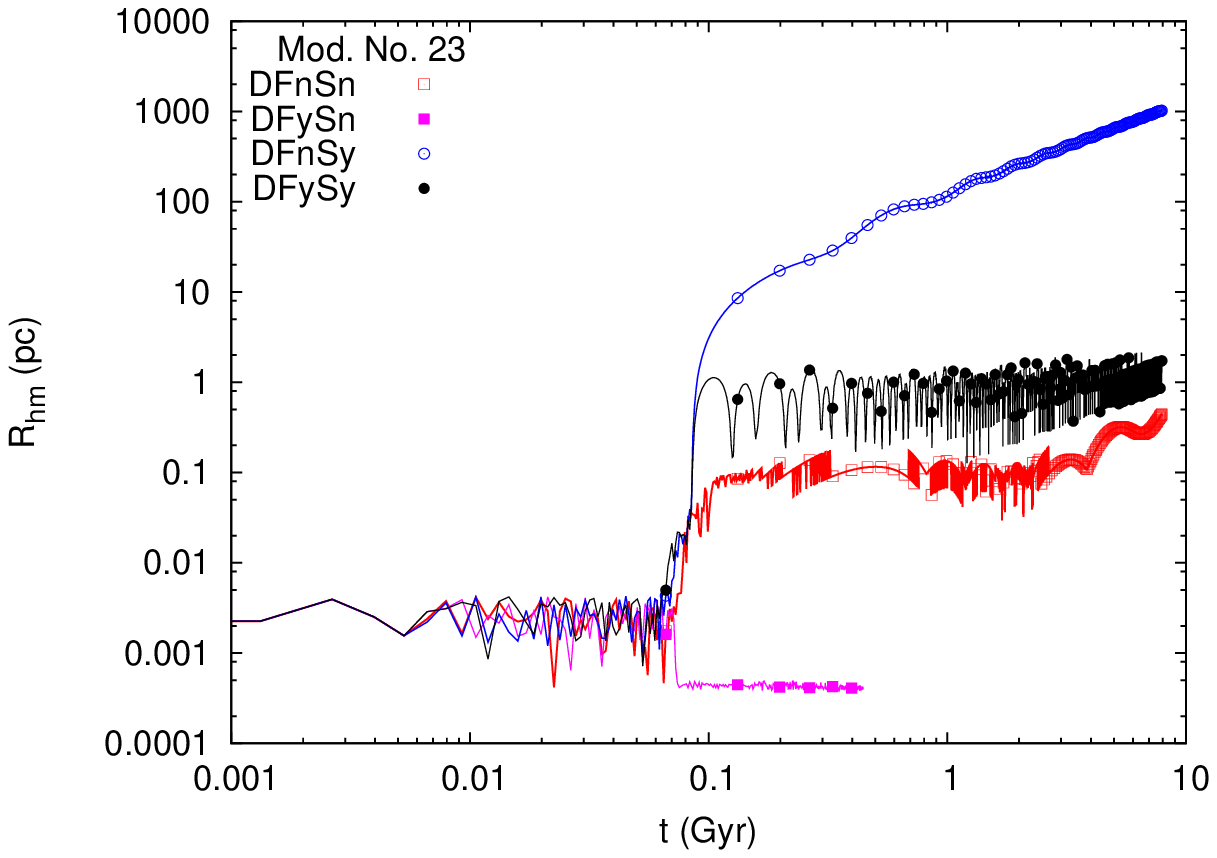}\\
\includegraphics[width=8cm]{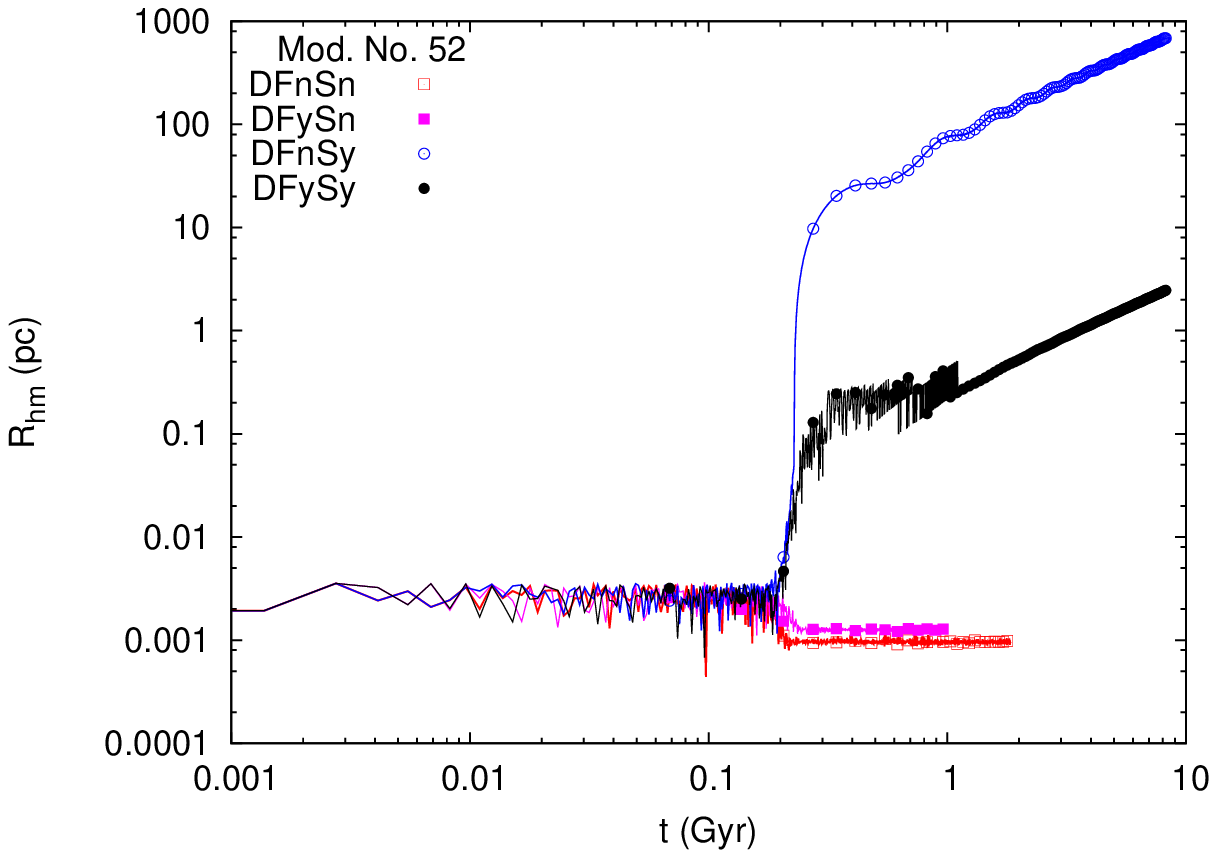}\\
\caption{The half-mass radius of the IMBHs population as a function of the time for four different simulations. In each panel, different curves refer to different groups, i.e. different physical processes taken into account.}
\label{hm}
\end{figure}

For each model in each group, we calculated the number of merger events, the mass of the merged IMBH and the time at which the event occurs.
Our models, carried out up to 13 Gyr, suggest that at most three IMBH can merge into the SMBH, while the other remain orbiting in its surrounding or are ejected away with speed up to 1000 km s$^{-1}$.
Figure \ref{hm} shows the time evolution of the half-mass radius $R_{\rm hm}$ of the IMBHs sample for four different runs, one for each models set. For the sake of comparison, each curve in each panel refers to the same set of initial conditions, but with different physical processes taken into account (dynamical friction on/off, zero/non-zero SMBH spin).

Note that here the half-mass radius is a parameter subjected to significant variations, due to the poor number of IMBHs in the sample. This causes the evident differences among the four configurations. For instance, the ejection of two IMBHs implies a $25\%$ reduction of the total IMBHs mass, which may result in a significant increase of $R_{\rm hm}$ calculated for the IMBHs sample. 

It is evident the crucial role of the initial conditions (different panels in figure \ref{hm}) and the physical features (different curves in each panel) in shaping the evolution of such a complex system, although it is composed only of 8 particles.
In some cases, dynamical friction tends to keep the IMBH population more concentrated, while the SMBH spin, when the dynamical friction is ``turned on'', determines a stronger contraction of the IMBH system, leading to a half-mass radius smaller than 0.05 pc.
In other models, however, dynamical friction causes stronger interactions between the IMBHs which, in turn, causes a rapid ejection of the lighter components, while this does not occur when it is not taken into account, due to the longer time-scales over which the 8-body system evolves. In any case, the chaotic nature of the few body problem, especially when the mass ratio  is large, makes poorly significant scaling correlations of results with large scale (average) properties of the motion environment.

Table \ref{multimer} outlines the number of mergers occurring between the IMBHs and the SMBH, specifying the fraction of cases in which the event involves the initial inner binary. Note that in several models the SMBH swallows two or even three IMBHs, leading the total number of mergers to exceed the total number of runs.
Despite a detailed investigation of the parameter space and the role played by different physical processes is beside the scope of this work, and despite the caveat about the stochasticity of few body dynamics which can overwhelm possible general trends, we note here that the inclusion of dynamical friction and a rotating (Kerr) SMBH seems to slightly decrease the merger probability, while is less trivial the effect that they have on the evolution of the inner binary. 
This is likely due to the fact that when dynamical friction is turned on, a larger number of IMBHs interact more closely in a complex way, thing that that may lead to the ejection of one or more components before it undergoes a merger. 

\begin{table}
\caption{Fraction of merger events in the multiple IMBHs simulations}
\begin{center}
\begin{tabular}{cccccc}
\hline
group name & $N_{mod}$ & $N_{mer}$ & $\nu_{mer}$ &$N_{inn}$ & $f_{inn}$ \\
 & & & & &$\%$ \\
\hline
DFnSn & 150 & 175 & 1.167&68 & 38.8\\
DFnSy & 163 & 185 & 1.135&66 & 35.7\\
DFySn & 73  & 95  & 1.301&29 & 30.5\\
DFySy & 88  & 111 & 1.261&44 & 39.6\\
\hline
\end{tabular}
\end{center}
\begin{tablenotes}
\item Column 1: group ID. Column 2: number of simulated models. Column 3: total number of merger events. Column 4: fractional number of mergers normalized to the total number of simulations in each group. Column 5-6: number of inner binary coalescence and their percentage over the total number of merger events.
\end{tablenotes}
\label{multimer}
\end{table}

\begin{figure*}
\includegraphics[width=8cm]{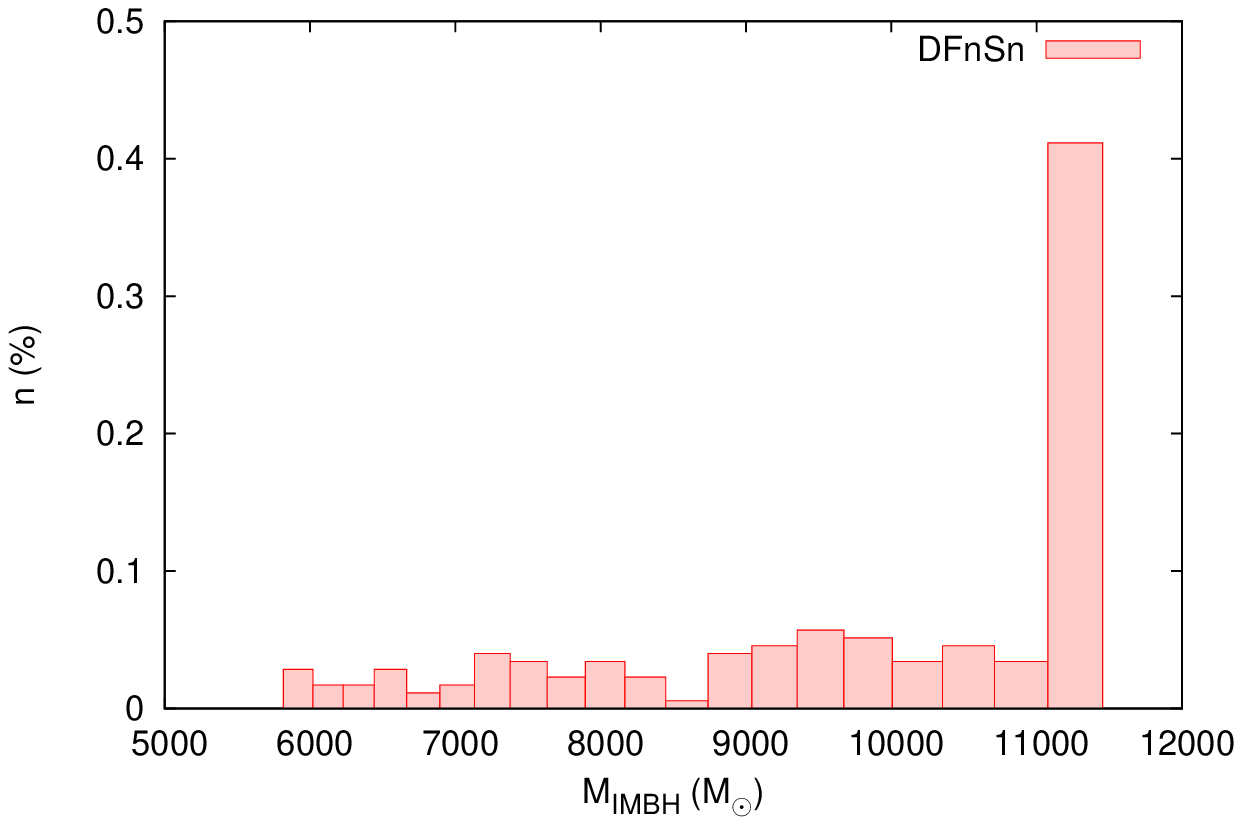}
\includegraphics[width=8cm]{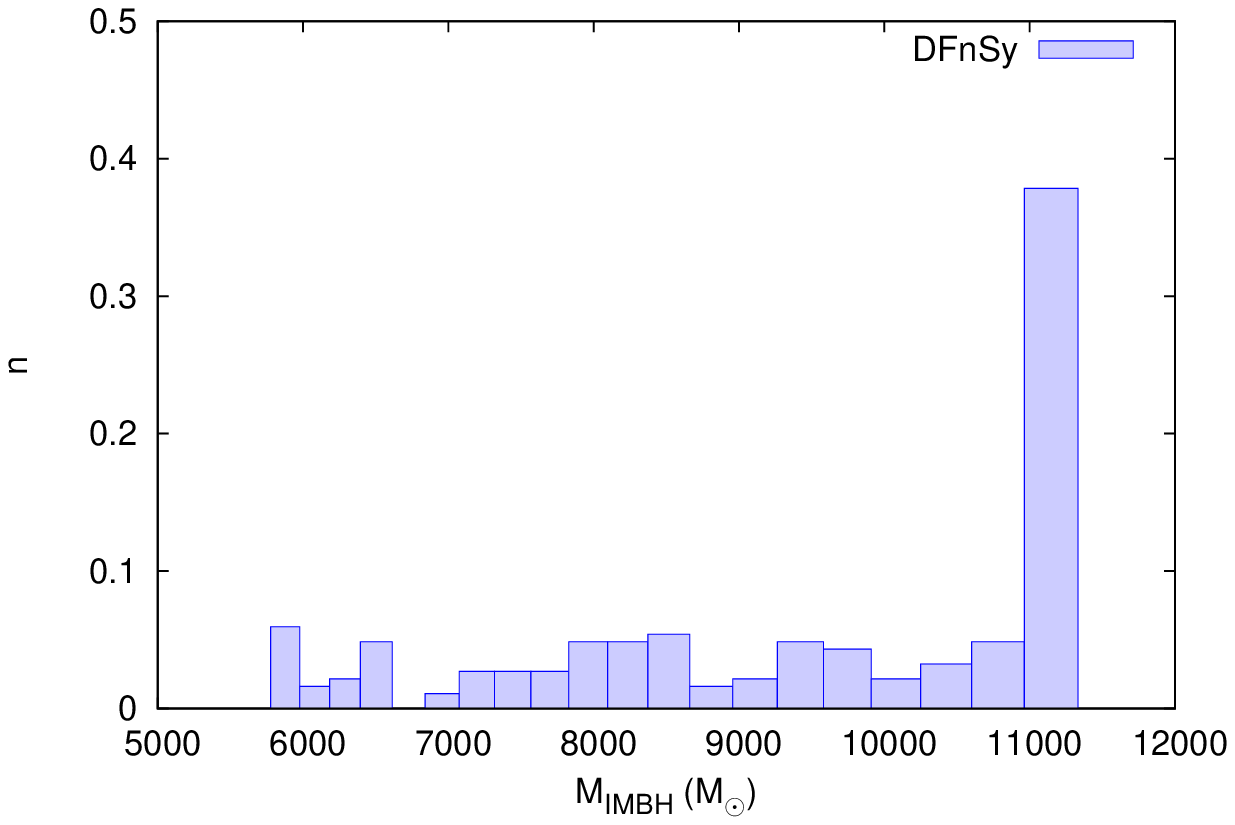}\\
\includegraphics[width=8cm]{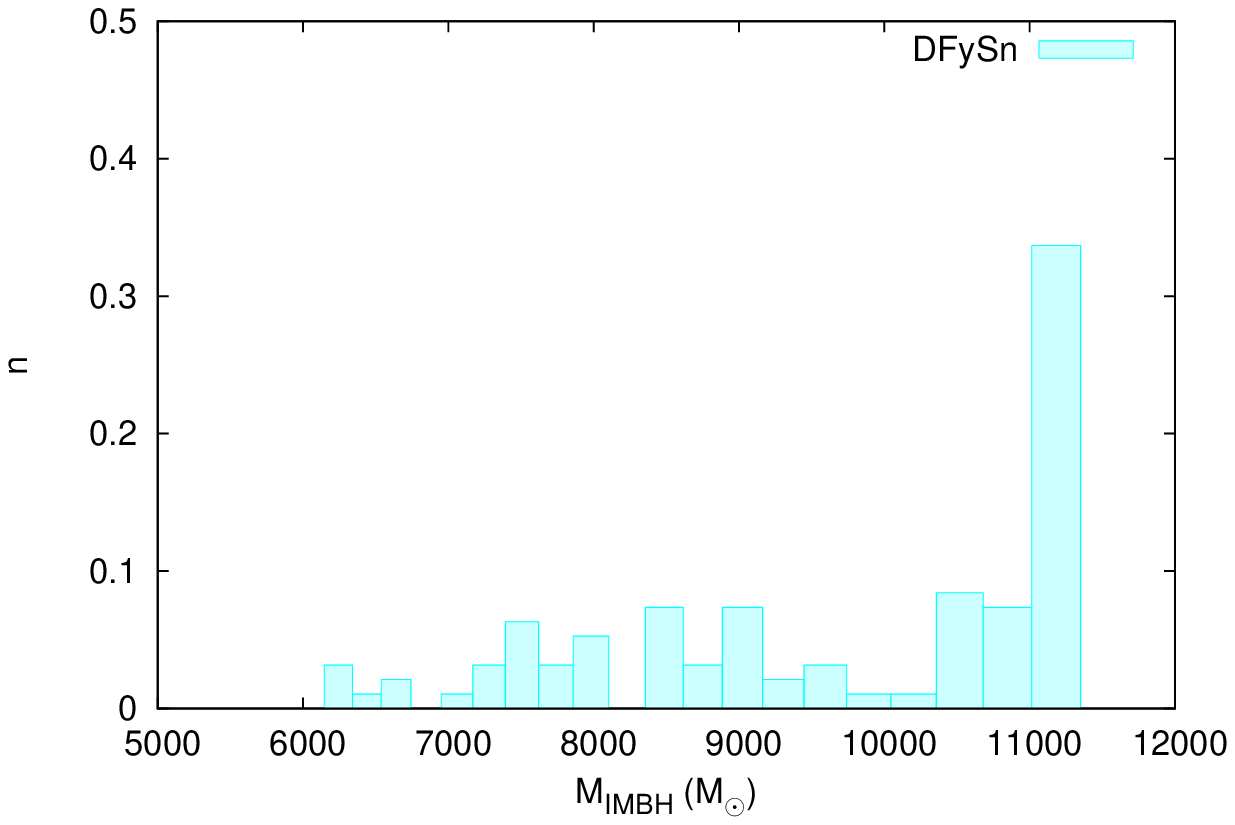}
\includegraphics[width=8cm]{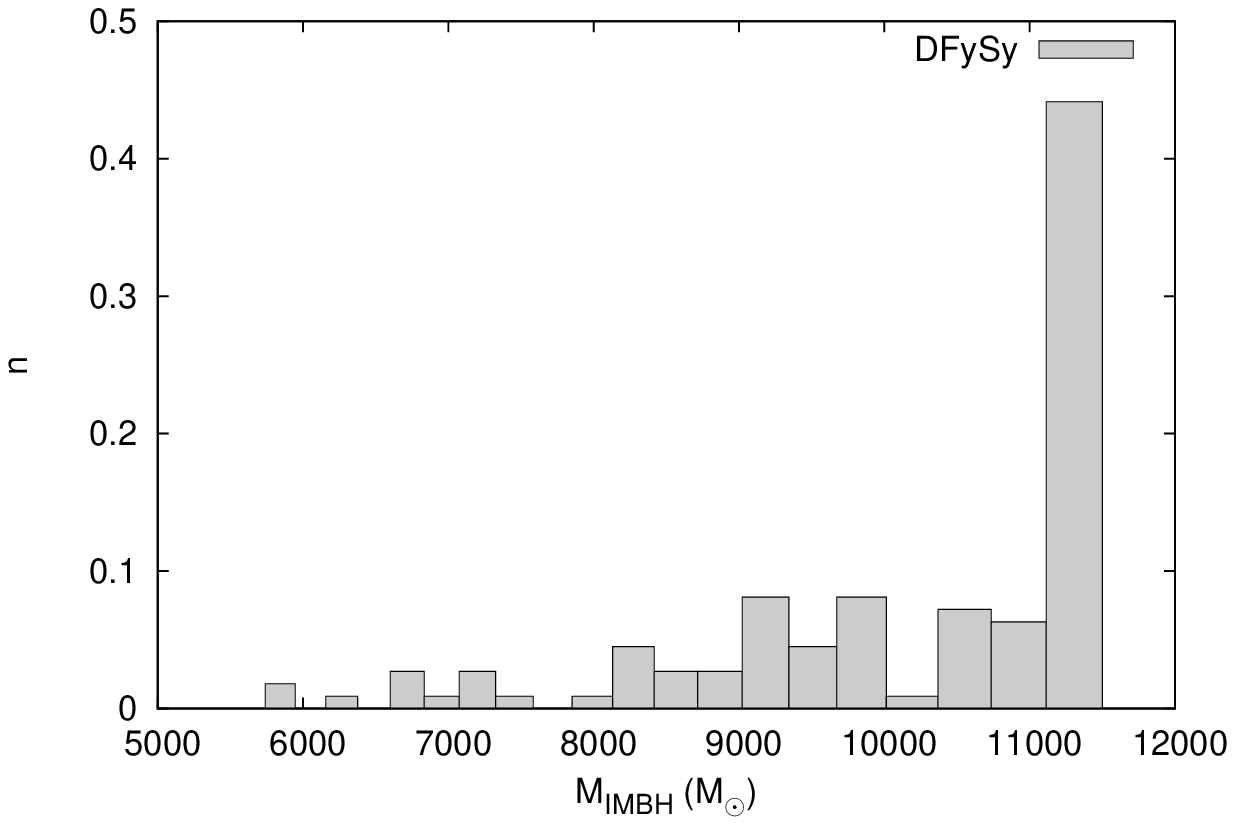}
\caption{Mass distribution of the merged IMBH in all the simulations groups investigated.}
\label{mhisto}
\end{figure*}

Figure \ref{mhisto} shows, for each group, the mass distribution of the merged IMBHs. Note that the distribution peaks in correspondence of the largest bin value, which contains the mass of the IMBH in the inner binary. The evident peak is clearly due to the fact that nearly $1/3$ of the mergers involve the initial inner binary.

\begin{figure*}
\includegraphics[width=8cm]{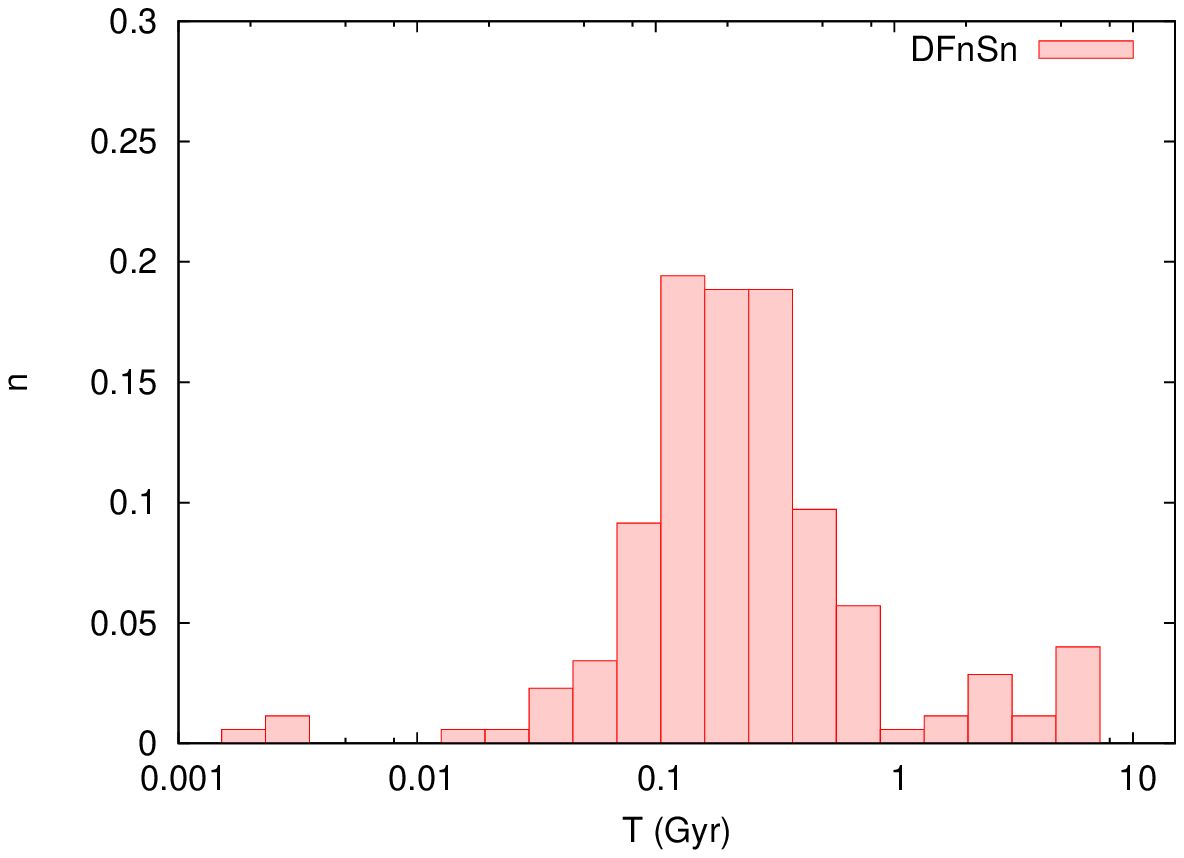}
\includegraphics[width=8cm]{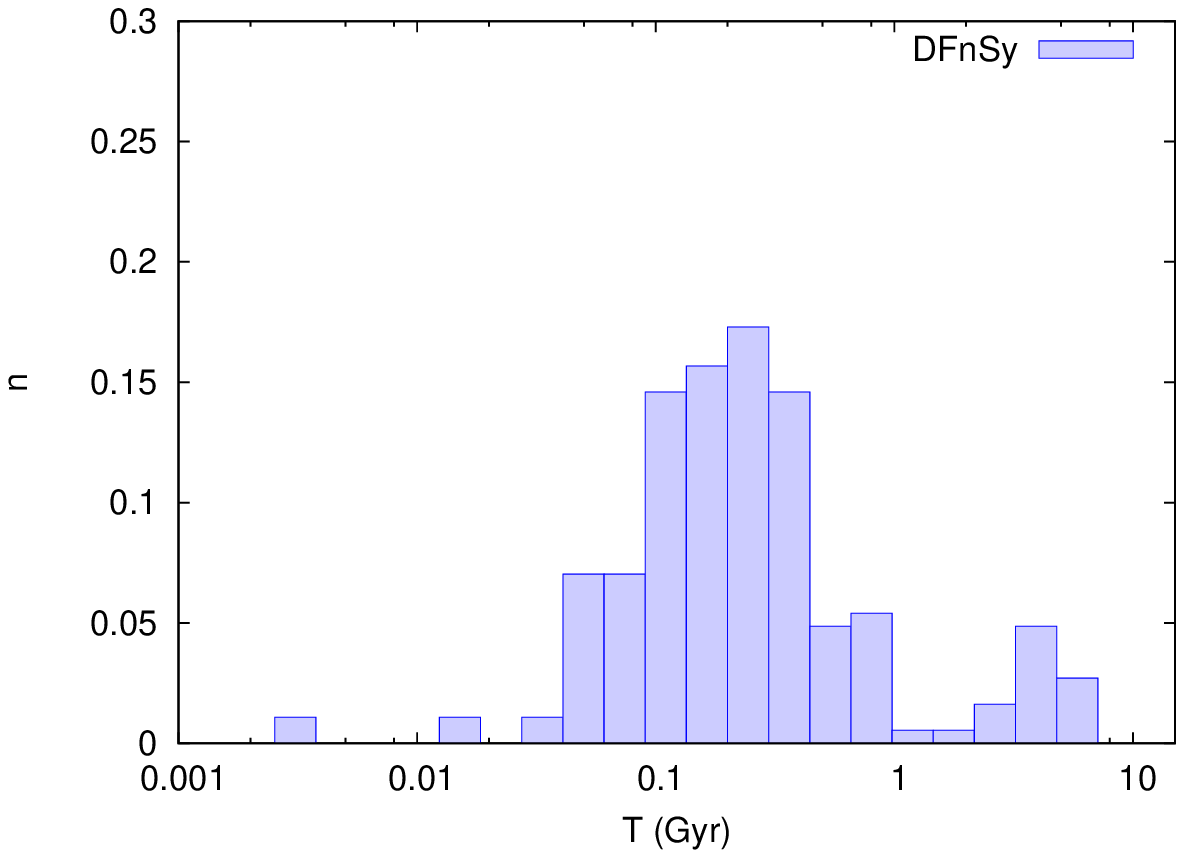}\\
\includegraphics[width=8cm]{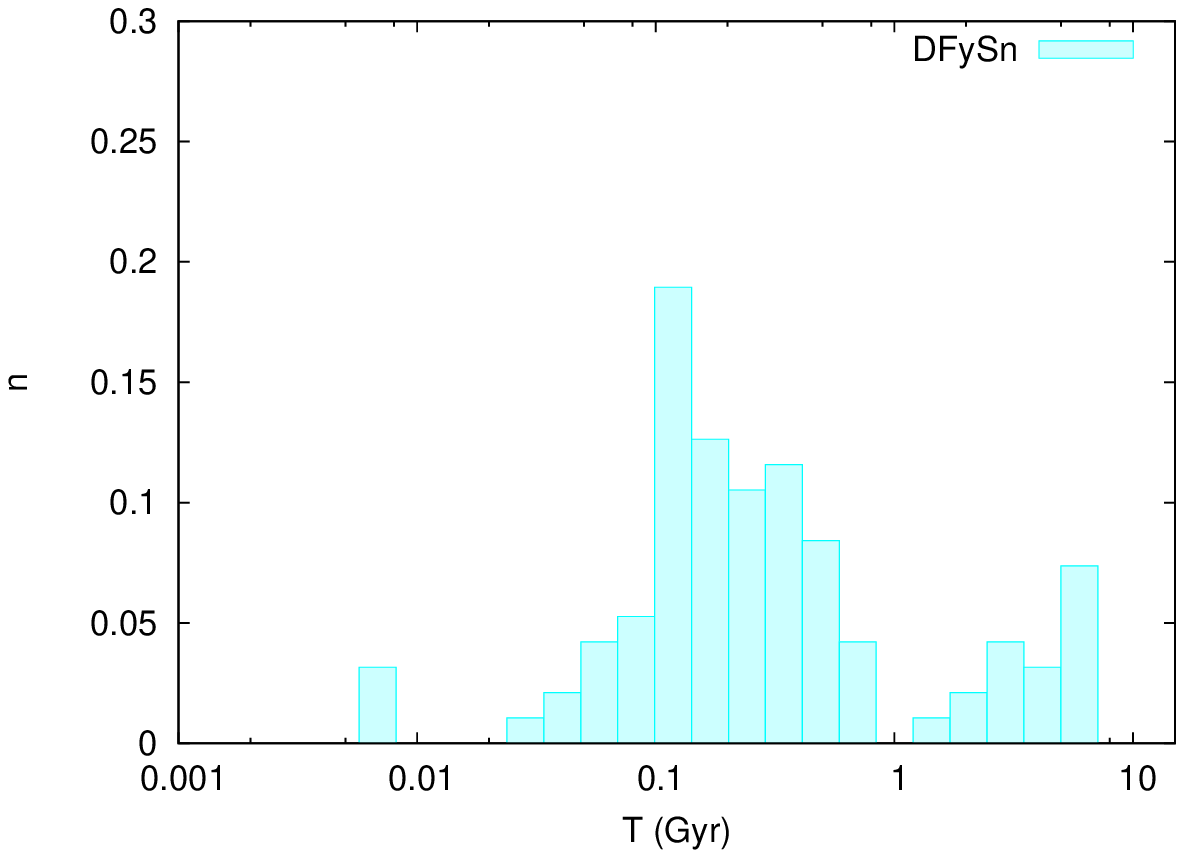}
\includegraphics[width=8cm]{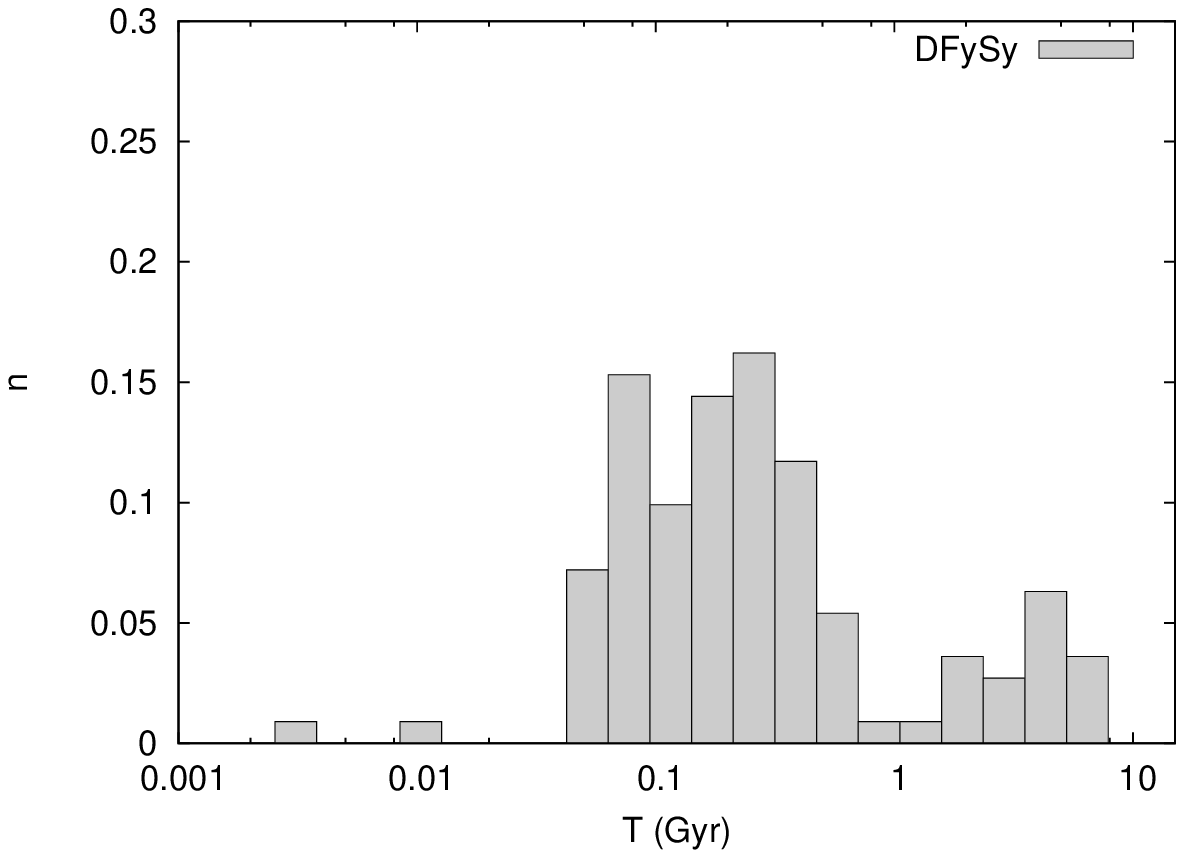}
\caption{Distribution of the time at which a merger occurred in all the models investigated.}
\label{thisto}
\end{figure*}

The time at which a merger occurs has a weak dependence on the model group considered, as shown in figure \ref{thisto}. Comparing groups DFnSn and DFySn, it is evident that a merger occurs earlier when dynamical friction is taken into account. When a spinning SMBH is considered (models DFnSy and DFySy) the merger occurs later, on average. In absence of dynamical friction and with a Schwarzschild SMBH, the merger time distribution peaks at $\sim 0.2$ Gyr. The peak slightly increases up to $0.3-0.5$ Gyr when a Kerr SMBH is considered and dynamical friction is turned off.

Our results show that when general relativity effects are coupled to dynamics, at least one IMBH-SMBH merger is unavoidable. 
This means that the presence of a small population of IMBHs around a central SMBH can drive several merger events within a Hubble time. The approximate rate at which these events might occur in a galaxy can be inferred using the $\nu$ parameter in table \ref{multimer} and the galaxy number density at redshift $z=0$ calculated in Section \ref{EMRI}: 

\begin{equation}
\Gamma_{\rm imbh} \simeq \frac{\nu}{10 {\rm Gyr}}n_g = 0.03 {\rm yr}^{-1}{\rm Gpc}^{-3}.
\label{probimbh}
\end{equation}

It is well known that after the merger the resulting SMBH can receive a recoil kick that can even eject it from the galactic nucleus.
According to \cite{Schnittman07}, the velocity kick imparted to the merged BHs is linked to the parameter $\eta = m_1 m_2 / (m_1+m_2)^2$. In our case, $\eta \sim 10^{-4}$, corresponding to a negligible kick on the final SMBH $v_{\rm kick}<50$ km s$^{-1}$ .
Even if we consider smaller masses for the SMBH, which imply higher $\eta$ values, the kick remains smaller than the galaxy central escape velocity. For instance, if we consider the Galactic SMBH mass ($4.5\times 10^6\Ms$), the kick remains below 100 km s$^{-1}$.

On another hand, a significant kick can be experienced by MBHs  with masses around $10^5\Ms$, expected to form in dwarf galaxies. Indeed, in such cases the recoil kick velocity ranges in between 100 - 300 km s$^{-1}$, large enough to result into an SMBH ejection. 

Using, again, the data available from the Illustris simulation, we found that galaxies with stellar masses in the range $10^8-10^9\Ms$, where dwarf galaxies lie, are characterized by a number density $n_{\rm dw}\sim 0.01$ Mpc$^{-3}$ at low redshift.
If we assume, quite arbitrarily, that only $P_{\rm dw}\sim 1\%$ of these objects hosted only one SMBH-IMBH merger event,$\nu_{\rm dw} = 1$ ,we can calculate the number density of SMBHs recoiled from dwarf galaxies by multiplying the dwarf galaxies number density by the SMBH-IMBH probability above (Eq. \ref{probimbh}),  obtaining
\begin{equation}
n_{\rm recoil} = P_{\rm dw} n_{\rm dw} \nu_{\rm dw} \simeq 10^5 {\rm Gpc}^{-3}.
\end{equation}

\bigskip

\subsection{The effect of SMBH-IMBH interactions on stellar BH binaries}
\label{binaries}

In this section we investigate the likelihood of a strong scattering of the SMBH with an IMBH around which is orbiting a pair of stellar BHs (a black hole binary, BHB). In order to investigate a wider range of IMBH masses, we varied its mass in the range $M_{\rm IMBH} = 10^3-10^6\Ms$.

\begin{figure}
\centering
\includegraphics[width=8cm]{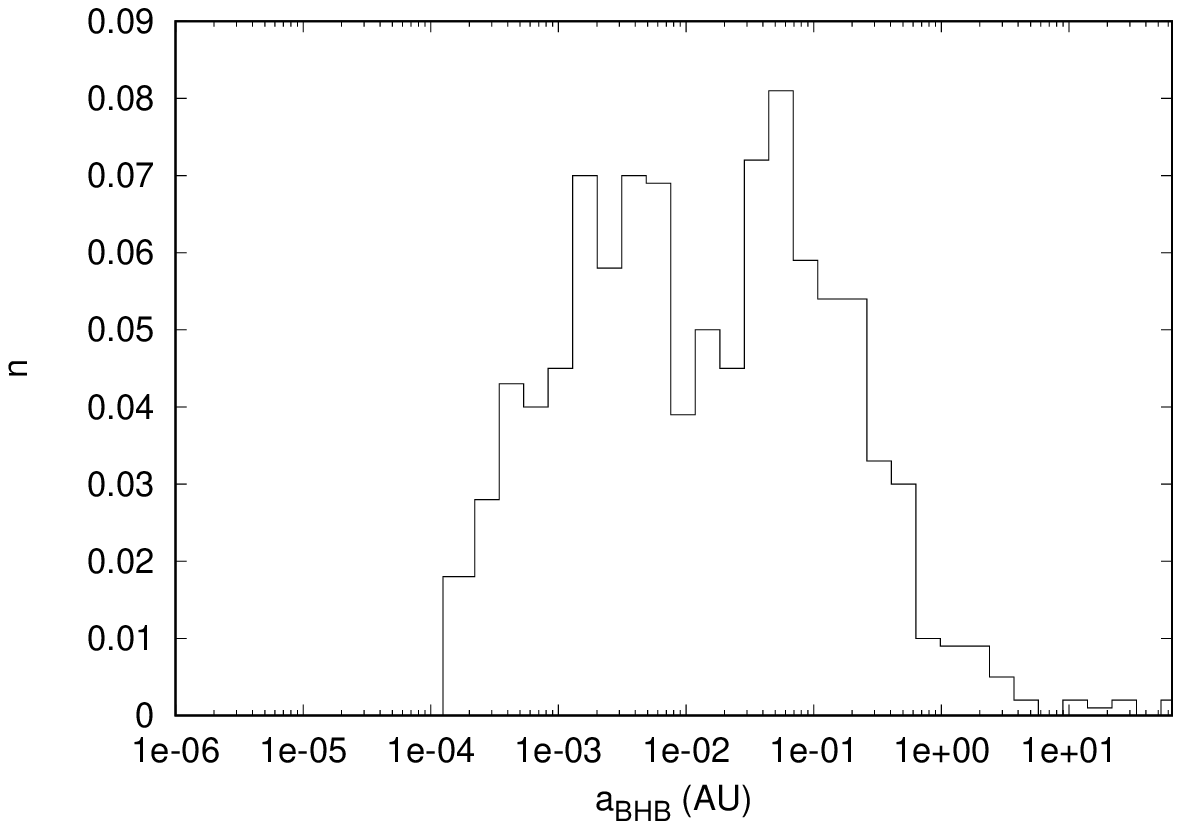}\\
\includegraphics[width=8cm]{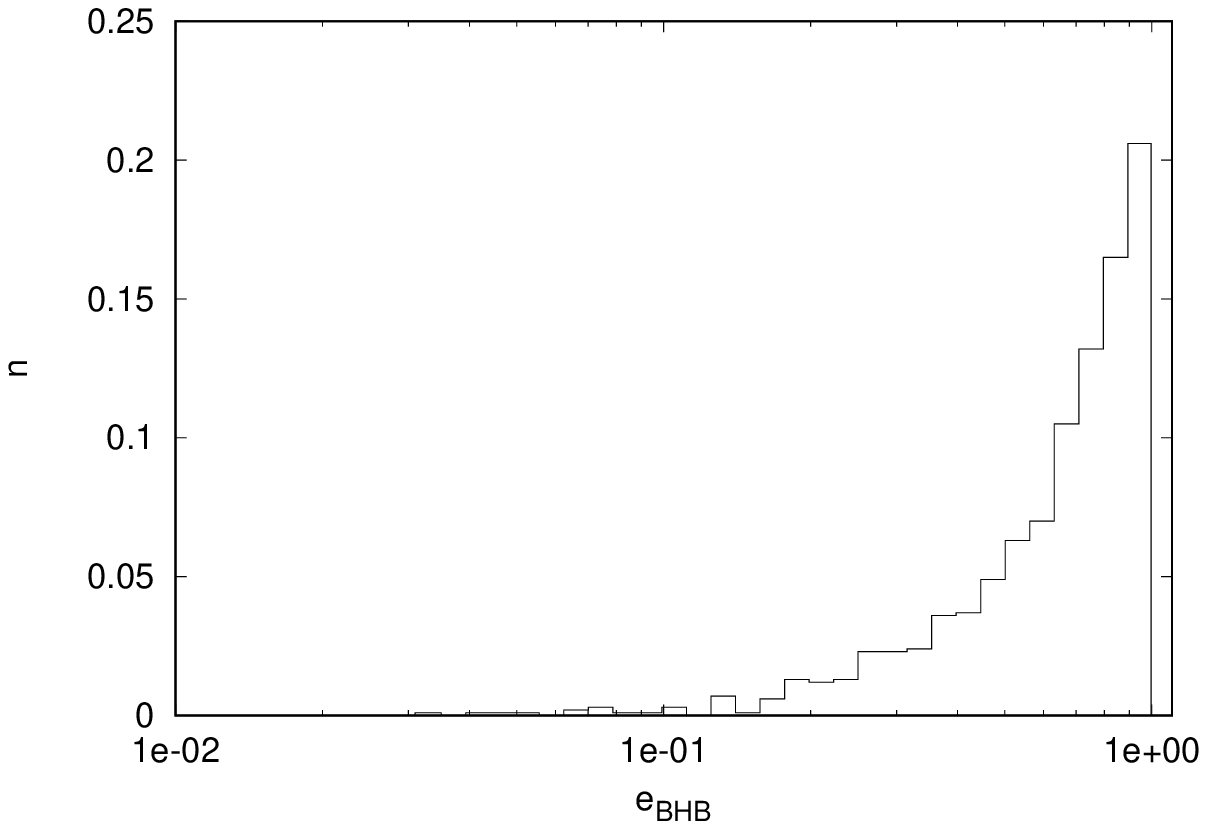}\\
\includegraphics[width=8cm]{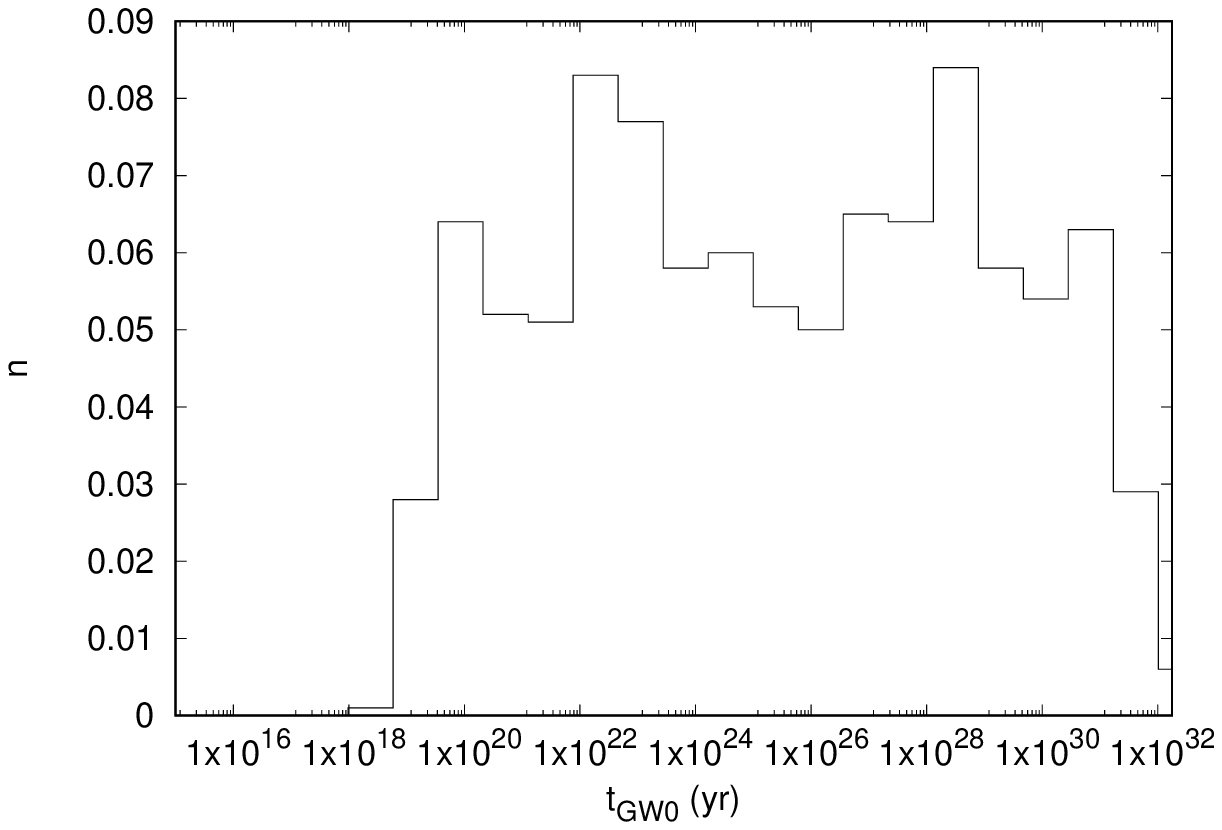}\\
\includegraphics[width=8cm]{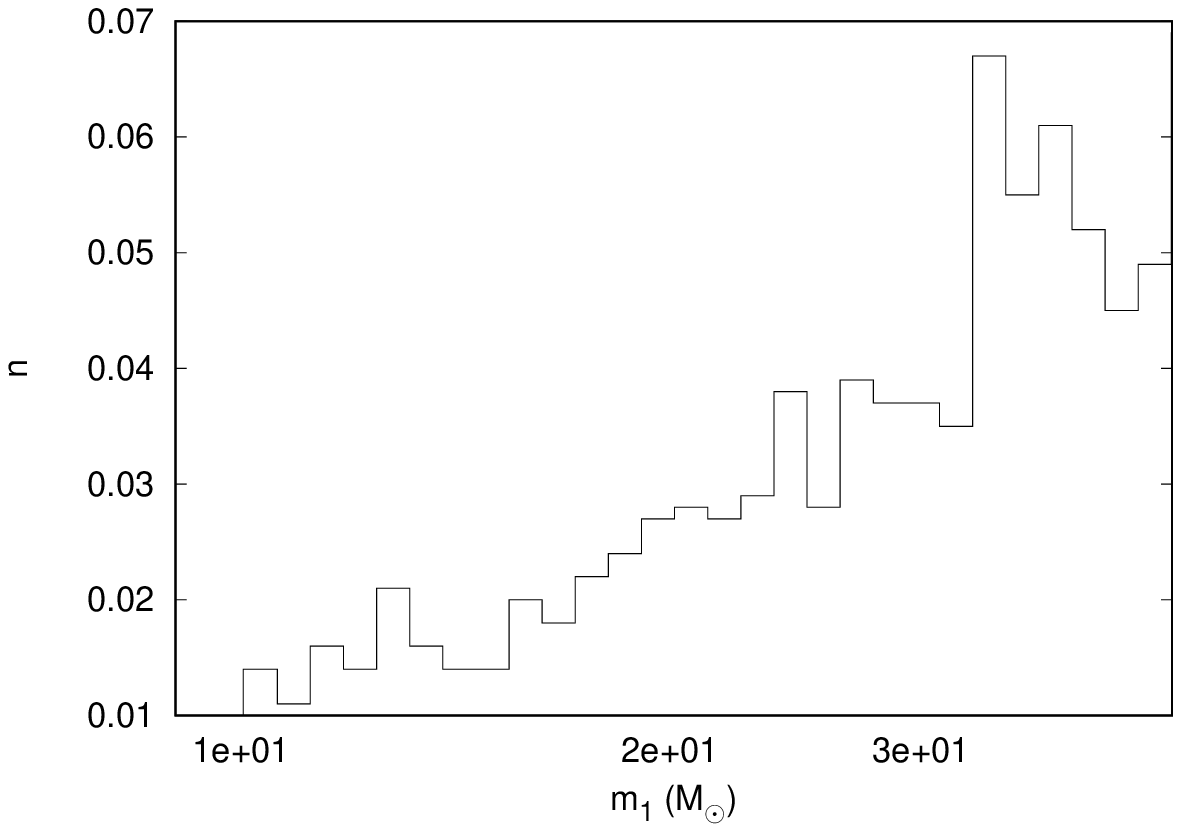}
\caption{From top to bottom: initial distribution of our stellar BHBs semi-major axis, eccentricity, GW time-scale and component masses. }
\label{bhbinit}
\end{figure}

The parameter space of such a problem is huge as it involves at least 10 degrees of freedom: the BHB properties (semi-major axis, eccentricity, mass of the components), the BHB orbit around the IMBH (semi-major axis, eccentricity, inclination), the outer IMBH and its orbit (semi-major axis, eccentricity, mass). 

We take advantage of \ARGdf to run 1000 different models in which all these quantities are varied. Figure \ref{bhbinit} shows the initial BHB properties in our sample. Note that the GW time in all the cases is much larger than a Hubble time.
The BHB semi-major axis $a_\bhb$ is selected between 20 AU and 0.1 pc in bins equally spaced in the decimal logarithm, while the BHB eccentricity distribution is picked from a thermal distribution \cite{jeans19}.

The triple system composed by the IMBH and the BHB moves around the SMBH following an orbit characterized by an initial pericentre $r_p = 0.1$ pc, while the orbital eccentricity is distributed between 0 and 1 according, again, to a thermal distribution. 
All the simulations have been carried up to $\sim 100$ orbits around the SMBH, i.e. about $2.6\times 10^6$ yr. 

Due to the complex interaction among the BHB, the IMBH and the SMBH, we found several different outcomes:

\begin{itemize}
\item in 647 cases out of 1000 the binary is disrupted;
\item in 3 cases the BHB merges before the triple (IMBH+BHB) accomplishes the first pericentre passage around the SMBH;
\item in 15 cases one of the BHB components is ejected, while the other pairs to the IMBH, forming a tight binary that coalesces within $10^8$ yr, according to the \cite{peters64} formula;
\item just in 1 case out of 1000 one of the BHB components merge with the SMBH.
\end{itemize}

All these outcomes may result into interesting consequences on the long-term evolution of the galactic nucleus. 

As said above, the strong acceleration imparted from the SMBH and the IMBH on the BHB leads to a high probability of BHB disruption, indicating that only $25\%$ of BHB can survive during the galactic nucleus assembly. 

We note that the models presented above represent an acceptable frame of approximation of the motion of a BHB around the core of a dense GC, although our choice to model an IMBH rather than the whole GC is dictated by the computational limitations imposed by the \ARGdf code, which is optimized to simulate only $N\lesssim 100$ particles. 

The lifetime for BHBs in GCs is, on average, long ($O(1~{\rm Gyr}$)) compared to the typical formation time of the galaxy nucleus ($O(0.1~{\rm Gyr}$)), thus implying the possibility that a BHB orbiting within a GC that approaches closely an SMBH can be disrupted and its components are left in the SMBH surroundings,and possibly captured as EMRI. 
As a consequence, the number of BHBs that can evolve within the growing nuclear cluster will be reduced significantly compared to the total BHB population in GCs. 
Formation of new BHBs through three-body scattering will be strongly suppressed due to the high velocity dispersion of stars in the central part of the galaxy. Indeed, the time-scale over which three-body interactions occur is given by 

\citep{lee95,antonini16} 
\begin{equation}
t_{3} \propto 4{\rm Gyr} \left(\frac{\sigma}{30{\rm km s^{-1}}}\right)^9.
\end{equation}

In our study, $\sigma \sim 100$ km s$^{-1}$, which implies $t_3 \simeq 2\times 10^5$ Gyr. 

Hence, our results suggest that the formation of new BHBs is strongly suppressed in galactic nuclei hosting an SMBH heavier than $10^8\Ms$, unless the binaries have formed before the SMBH growth or have been delivered to the galactic centre by orbitally segregated star clusters.

In a few models, among the remaining $25\%$ in which the BHB survives to the strong interaction with the IMBH and the SMBH, we found that the complex dynamics leads either to the BHB merger or { to the formation of a tight IMBH-BH pair that eventually merge on a time scale $\sim 10^8$ yr. The latter kind of binary systems is usually referred to as intermediate mass ratio inspiral (IMRIs) and is expected to be a kind of GW source particularly interesting due to its likely observability with both ground- and space-based detectors \citep{amaro12,seoane18}}.

In the following, we quantify the rate at which these rare mergers occurs in massive elliptical galaxies. 
First of all, we assume that the GCs total mass is $f_{\rm gc} = 0.01$ times the host galaxy mass, and that only $f_{\rm dec} \sim 11\%$ of the GC population segregates into the galactic nucleus within a Hubble time, as recently suggested by \citep{belczinski17}. Then, assuming a thermal distribution for $e_\gc$, the fraction of GCs having orbits more eccentric than $0.5$ is $f_{\rm ecc} = 0.75$. As suggested by \cite{Giersz15}, the fraction of GCs hosting an IMBH is relatively low, $f_\ibh = 0.2$. Therefore, the number of infalling clusters in an elliptical galaxy with mass $M_g \simeq 10^{11}\Ms$ and assuming a typical GC mass $M_{\rm gc}\simeq 1.5\times 10^6\Ms$ \citep{harris14,webb15}
will be

\begin{equation}
n_{\rm dec} = f_\ibh f_{\rm gc}f_{\rm dec}f_{\rm ecc}\frac{M_g}{M_{\rm gc}} = 11.
\label{EQndec}
\end{equation}

According to a simple stellar population characterised by a \cite{kroupa01} mass distribution, the fraction of stars sufficiently massive to turn into BHs, i.e. with ZAMS mass $18\Ms$, is $f_{\rm bh} = 2\times 10^{-3}$. 
Assuming that only a fraction $\eta_{\rm bh}=1-10\%$ of the BHs population go into a binary, the number of BHBs expected, on average, to form in a GC is given by

\begin{equation}
n_{\rm bh} = \eta_{\rm bh} f_{\rm bh} \frac{M_\gc}{m_*} 
\label{EQnbh}
\end{equation}

which ranges in the interval $= 30-300$, in dependence on the value of $\eta_{\rm bh}$.

The deposit of BHBs into the galactic centre is characterised by the dynamical friction time-scale of the progenitor GC, 
which can be written as  \citep{ASCD14a,ASCD15He}

\begin{equation}
\tau_{\rm df}(m_*) \simeq 0.3 {\rm Myr} \, g(e_{\rm GC},\gamma_g)\left(\frac{M_{\rm GC}}{M_g}\right)^{-0.67},
\label{tdf}
\end{equation}
where $g(e_{\rm GC},\gamma_g)\simeq 1-4$ is a smooth function of the GC eccentricity and the galaxy inner density slope.

On another hand, the time-scale for BHs segregation and pairing will be of the order of the mass-segregation time-scale, although 
its precise value depends on specific orbit and GC characteristics \citep{gaburov08,downing10,askar17}. As discussed in \cite{AS16}, Eq. \ref{tdf} conveniently manipulated can be used to crudely estimate the BHs mass segregation time-scale in their progenitor GC. 

\begin{figure}
\centering 
\includegraphics[width=8cm]{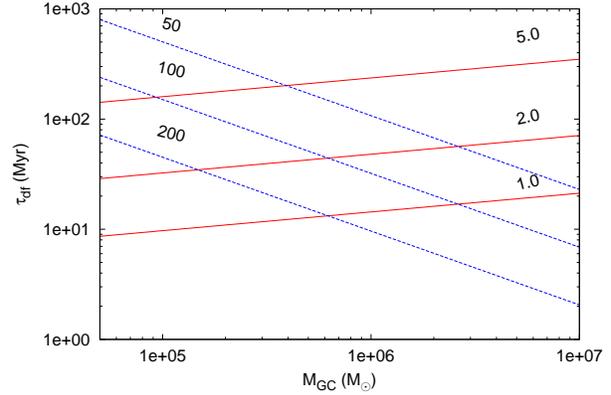}
\caption{Dynamical friction time as a function of the GC mass. Red straight lines refer to  a $25\Ms$ stellar BH in the hypothesis it moves initially at 1, 2 or 5 times the GC core radius. Blue dashed lines refer to the orbital decay time of GCs in the galaxy and are labelled by different values of the GC apocentre.}
\label{tdfcmp}
\end{figure}

Figure \ref{tdfcmp} shows a comparison between $\tau_{\rm df}$ and $\tau_{\rm dfBH}$ at varying BH orbital properties and GC mass and orbital pericentre. 
If $\tau_{\rm df}$ does not exceed severely $\tau_{\rm dfBH}$, then the formation of BHBs within the host cluster is a process still working when the host cluster impacts the SMBH.
This, in turn, implies the possibility that a consistent population of BHBs would be still present in the GC core at the moment of its close encounter with the central SMBH.
For our calculations, we assume a typical value of $t_{\rm mer} = t_{\rm df} = 10^8$ yr, if not otherwise specified. 

Plugging equations \ref{EQndec}-\ref{tdfcmp} in Eq. \ref{EQgamma}, having assumed $t_{\rm mer} = \tau_{\rm df}$,  we calculate the merger rates for all the different channels found in our runs.

As listed above, in $0.3\%$ of the models the BHB merged before the IMBH+BHB system approached the pericentre. 

Hence, the rate of BHB merger events mediated by an IMBH in GCs impacting a heavy SMBH in an elliptical galaxy is 
\begin{equation}
\Gamma_{\rm BHB,IMBH} = 2 ~ {\rm yr}^{-1} {\rm Gpc}^{-3},
\end{equation}
where we assumed that only $1\%$ of the BHs are in binary systems around the IMBH ($n_{\rm bh}=30$, see Eq. \ref{etabh}), a value compatible with the merger rate for NSCs without any central SMBH \citep{antonini16}, and slightly smaller than for GCs \citep{askar17,rodriguez15} or  for young clusters \citep{banerjee16,banerjee18,mapelli16}. 
In principle, since the BHB merger here is only mediated by the presence of the IMBH, while it is not related to the GC infall, we could assume $f_{\rm dec}f_{\rm ecc} = 1$, so to include all the GCs having a central IMBH. 
In this optimistic case, the number of GCs would increase by an order of magnitude, being given simply by $n_{\rm dec} = f_\ibh f_{\rm gc}M_g/M_{\rm gc}=130$, leading to $\Gamma_{\rm BHB,IMBH} = 8 ~{\rm yr}^{-1} {\rm Gpc}^{-3}$.

On the other hand, in $15\%$ of { our models the IMBH gravitational pull causes the BHB disruption with one of the components being captured by the IMBH and yielding to the formation of an intermediate-mass ratio inspiral, which eventually merges in less than $10^8$ yr. }

These events can be potentially detected by the LIGO/VIRGO experiments at design sensitivity, although currently is not possible to detect GWs coming from an IMRI, and by the next generation of GWs detectors, such as LISA, TianQin, Taiji or the Einstein Telescope \citep{punturo10}. { As recently suggested by \cite{seoane18}, IMRIs can successfully be detected jointly with ground- and space-based observatories, and will likely constitute a significant step-forward to a more complete understanding of how BH pairs form and evolve in the Universe.}

This specific IMBH-BH merger rate is given by:

\begin{equation}
\label{etabh}
\Gamma_{{\rm IMRI}} = 9.5 {\rm yr}^{-1} {\rm Gpc}^{-3},
\end{equation}
compatible with similar estimates in literature \citep{fragione17c}.

\subsection{Stellar BH binaries left in the galactic centre.}

In the case in which the IMBH formation failed, the infalling GCs are expected to deliver around the galactic SMBH their BHBs. The motion of BHBs in the field of an SMBH can boost the BHB merger rate \citep{antonini12,hoang18}. 

In order to test this scenario, we ran 1000 simulations with ARGdf of a single BHB moving around the SMBH. As it was shown by \cite{ASCD17}, the GC debris likely distributes in a disky configuration around the SMBH, extending up to a few pc. 

We selected the BHB initial semi-major axis $a_\bhb = 10^{-4}-0.01$ pc, using bins equally spaced in the logarithm, while $e_\bhb$ is selected  according to a thermal distribution. The BHB components are assumed to have masses randomly chosen between $10$ and $50\Ms$.
In all cases studied the BHB have initial GW time-scales, estimated with the \cite{peters64} formula, larger than 10 Gyr. 
The orbital semi-major axis of the BHB around the SMBH, is drawn in the range $0.1 - 25$ pc, and also in this case we use a thermal distribution to distribute eccentricities.
All the simulations are run up to $t = 2.5$ Myr, that is about 3 revolution periods around the SMBH, time short enough to make our neglecting of the surrounding stellar field a reliable choice.

We found that, over this time-scale, the BHB merges in $5.2\%$ of the cases. This is due to the nature of Kozai-Lidov oscillations, whose intrinsic time scale 
\begin{equation}
t_{\rm KL} \propto \frac{2}{3\pi}\left(\frac{GM_{\rm SMBH}}{a^3}\right)\left(\frac{GM_\bhb}{a_\bhb^3}\right)^{-1/2},
\end{equation}
with $a$ the semi-major axis of the BHB with respect to the SMBH.

As a conclusion, promptly after the GC deposited its debris around the SMBH, the combined action of the tidal field of the stellar environment and of the SMBH drives the BHB toward a rapid coalescence. 

Following the same treatment described in the last section, we can calculate the BHB merger rate in the infalling scenario for massive elliptical galaxies as:

\begin{equation}
\Gamma_{BHB,SMBH} = n_{\rm dec} n_{\rm BHB} n_g t_{\rm mer}^{-1} = 1 ~{\rm yr}^{-1} {\rm Gpc}^{-3},
\end{equation}

where we assumed that only $1\%$ of all the BHs population formed a BHB with a sufficiently long life time to be delivered into the galactic centre, although this likely represents a lower limit.
Hence, the delivery of BHBs from massive GCs formed within the galactic nucleus and rapidly centrally segregated, on a time-scale $\sim 0.1$ Gyr, can contribute significantly to the GW emission from these regions.

As noted by \cite{Chen17}, the GW signals coming from a BHB merger revolving around an SMBH can possibly suffer a shift in frequency larger than what calculated in standard GW astronomy. 
The observables that can be extracted directly from GWs signals are the strain $h$, the frequency $f$ and its variation $\dot{f}$. 
Using these quantities we can infer the BHB chirp mass in the observer rest-frame
\begin{equation}
\mathcal{M}_o = \left(\frac{5f^{-11/3}\dot{f}}{96\pi^{8/3}}\right)^{3/5},
\label{chirp}
\end{equation}
and its distance 
\begin{equation}
d_o = \frac{4\mathcal{M}_o}{h}\left(\pi f \mathcal{M}_o\right)^{2/3}.
\label{lumdis}
\end{equation}

Since redshift, by definition, reduces the signal frequency $f$ by a factor $(1+z)^{-1}$ and its derivative $\dot{f}$ by a factor $(1+z)^{-2}$, it is trivial to show that the chirp mass in the observer rest-frame, namely $\mathcal{M}_o$, will be related to the actual value through a simple relation
\begin{equation}
\mathcal{M}_o = \mathcal{M}(1+z).
\end{equation}
Here, the intrinsic chirp mass $\mathcal{M}$ is a combination of the binary masses, $\mathcal{M}=(m_1m_2)^{3/5}/(m1+m_2)^{1/5}$, thus implying that the measured chirp mass carries information about the merging binary components.

The $(1+z)$ term must account for all the physical processes that can shift the GW frequency. For instance, the Universe accelerated expansion causes a signal redshift $z_{\rm C}$ that increases with the comoving distance $d_C$. This term is what is the only included in the standard procedure followed in GW data analysis, namely $\mathcal{M}_o = \mathcal{M}(1+z_C)$.
However, if the BHB moves around an SMBH, the GW signal emitted during the merger phases can suffer a Doppler shift $z_D$ that, according to special relativity, can be written in terms of the source velocity through the $\beta_D = v/c$ parameter 
\begin{equation}
1+z_{\rm D} = \gamma_D (1+\beta_D),
\label{dpplr}
\end{equation}
where $\gamma_D = (1-\beta_D^2)^{-1/2}$ is the Lorentz factor.

A further effect is due to gravitational redshift $z_G$, which develops in the case the BHBs semi-major axis is comparable to a few SMBH Schwarzschild radii \citep{Chen17}. However, this effect is negligible in our models and not included in the following calculations. 

Combining the three effects, from equations \ref{chirp}-\ref{lumdis}  it is possible to connect the measured and intrinsic chirp mass and distance to the redshift as 
\begin{align}
\mathcal{M}_o &= \mathcal{M}(1+z_C)(1+z_D)(1+z_G).\\
d_o &= d_C(1+z_C)(1+z_D)(1+z_G).
\end{align}

In our simulations, the BHBs delivered by spiralling GCs to the galactic centre have semi-major axis in the range $a = 0.1-10$ pc and eccentricities $e>0.5$, on average. Note that here we are referring to the BHB centre-of-mass orbit around the SMBH. During the passage at pericentre, the $\beta_D$ factor can be written as
\begin{equation}
\beta_D^2 = \frac{r_{\rm S}}{2a}\frac{1+e}{1-e},
\end{equation}
where we used the Schwarzschild radius, $r_S$, to replace the speed of light $c^2 = 2GM_{\rm SMBH}/r_{\rm S}$.

\begin{figure}
\centering
\includegraphics[width=8cm]{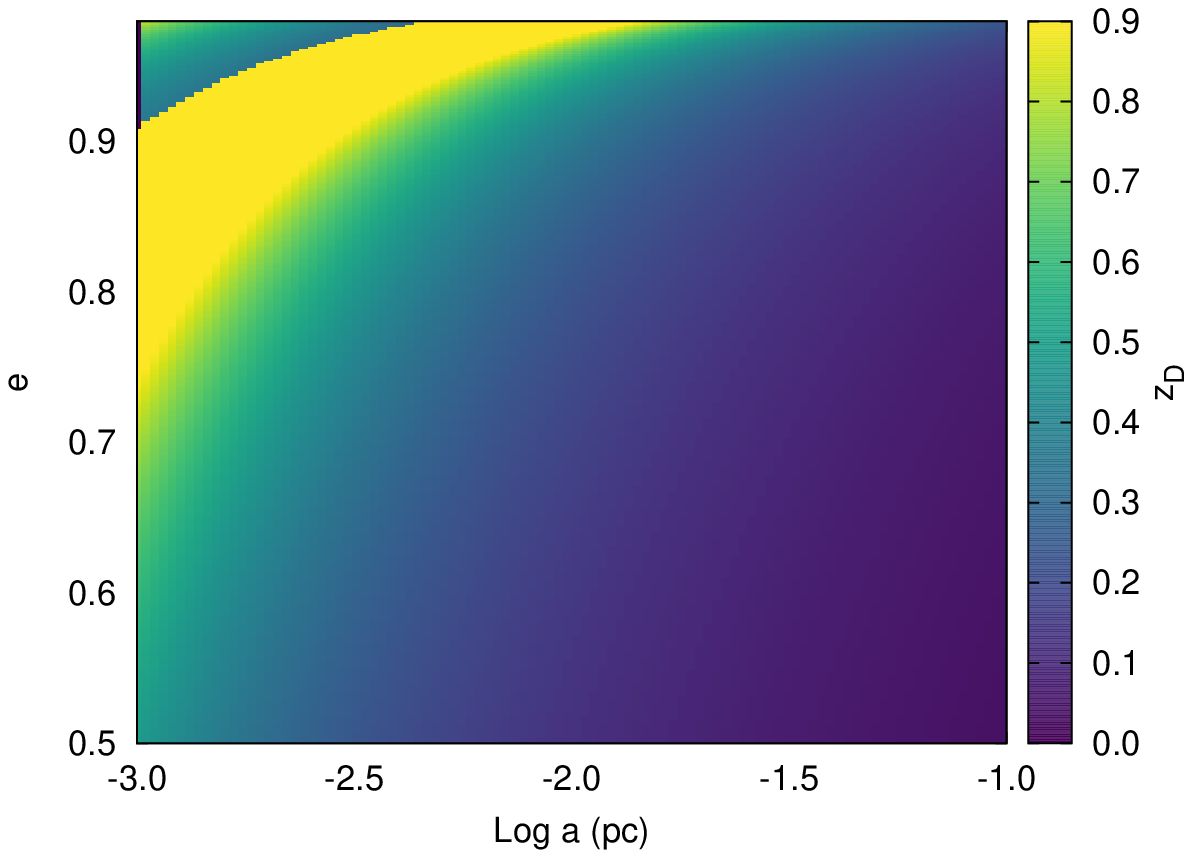}
\includegraphics[width=8cm]{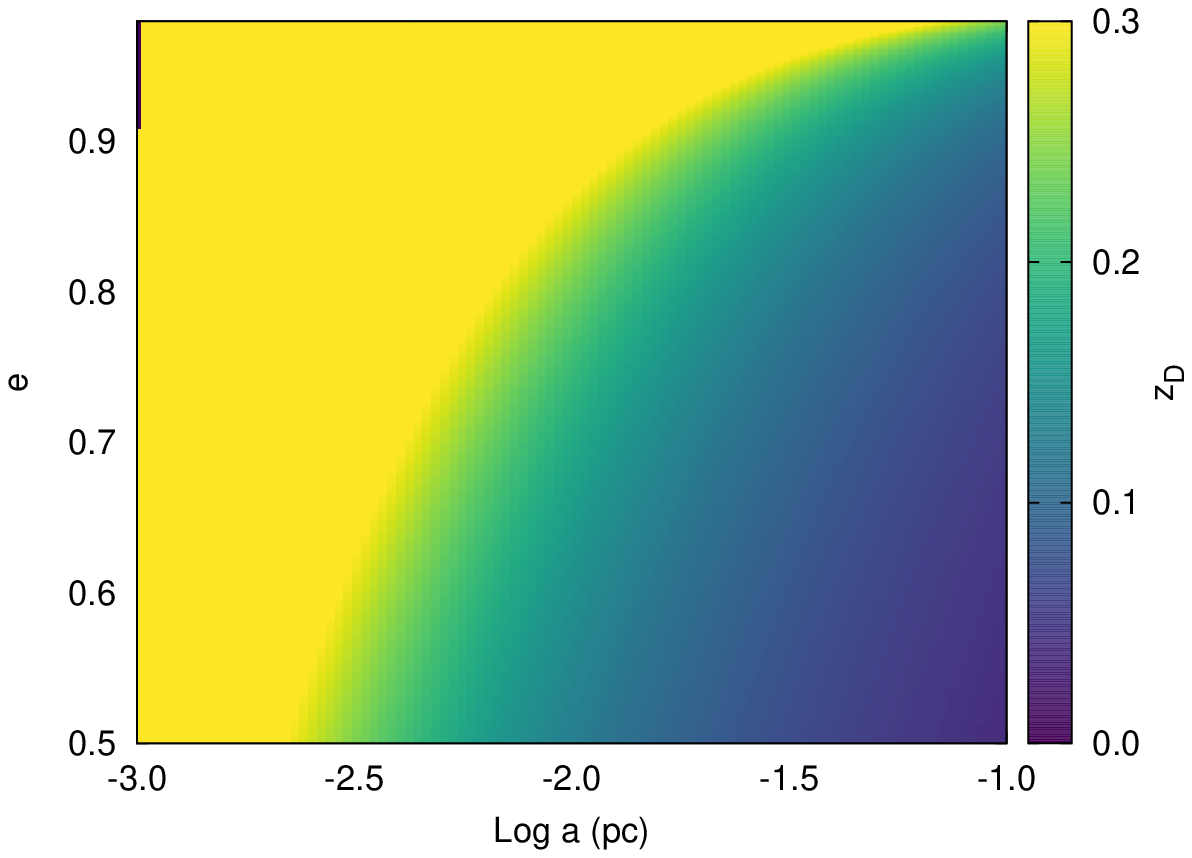}
\caption{Doppler shift as a function of the orbital semi-major axis and eccentricities for typical BHB orbits in our models, assuming an SMBH with mass $10^8\Ms$ (top panel), or $10^9\Ms$ (bottom panel). Note that the orbital parameters here refer to the BHB centre-of-mass motion around the SMBH.}
\label{redsh}
\end{figure}

Figure \ref{redsh} shows how the Doppler shift $z_{\rm D}$ varies for typical values of $a$ and $e$ and assuming an SMBH mass $10^8-10^9 \Ms$. 
{ It must be noted that equation \ref{dpplr} is valid as long as the line of sight lies on the orbital plane of the source, thus representing an upper limit of the Doppler shift induced by the motion around the SMBH on the GW signal.}
Assuming an SMBH mass $M_{\rm SMBH} = 10^8\Ms$ as in our direct N-body simulations, our calculations suggest that a non-negligible Doppler shift is already found at semi-major axis values $a<0.1$ pc, provided that the BHB orbital eccentricity is moderately large $e>0.8$. The shift increases considerably if the binaries move on tighter orbits ($a<0.01$ pc), even for moderate values of the eccentricity $e\gtrsim 0.5$. For these typical values, the shift amplitude is generally small, $z_{\rm D} < 0.2$.
Nevertheless, a non-null Doppler shift has profound implications on the typical BHB quantities calculated through GWs detection.
Mass segregation can shrink the BHB orbit, leading it in a regime where the redshift term can increase significantly, i.e. at semi-major axis values $a<10^{-3}$ pc.

If some of the GWs detected by LIGO originated from a BHB merger around an SMBH, the intrinsic chirp mass measured from the signal might need to be rescaled by a factor $(1+z_D)^{-1}$. Note that for nearly equal-mass BHBs the same factor applies to the component masses. 

Another way to look at this effect is that a BHB orbiting around an SMBH would appear heavier than the same binary evolving in a more ``quiet'' environment, like in the galactic field, where the Doppler shift is negligible.

In all the models investigated here the Doppler shift correction is generally small, characterised by an associated variation on the intrinsic chirp mass of $\sim 10-25\%$. 
This variation can be much larger for heavier SMBHs, which can lead to $z_D>0.5-0.7$ for moderate semi-major axis ($a\sim 0.01$ pc) and eccentricities ($e\gtrsim 0.5$), as shown in the bottom panel of figure \ref{redsh}.

\section{Conclusions}
\label{conclusions}

In this paper we investigate various channels of GWs emission emitted from the central region of a massive elliptical galaxy hosting a SMBH with mass $10^8\Ms$. To do this, we used the \ARGdf  code, an updated version of the \ARCHAIN code developed by \cite{mikkola08}. The code, as its progenitor, implements the algorithmic regularization treatment for close encounters and post-Newtonian terms up to order $2.5$ 
to account for general relativity effects. Our updates consist in the addition of a dynamical friction term, treated as an extension of \citet{Cha43I} theory, and the galactic field as an external potential. 

We tested results of \ARGdf against direct N-body modelling, performed with \HiGPUs \citep{Spera}, integrating the motion of a few massive bodies travelling in a sea of 131k particles, finding a very good agreement. 

Using our few-body code, we focused the attention on the role played by GC-SMBH interaction in the development of GWs. We investigated whether {\it disrupted} GCs can deliver toward the galactic centre an IMBH or several stellar BHBs, thus enriching the population of BHs existing in the inner galactic regions. In order to obtain reliable initial conditions for our study, we took advantage of the large amount of data provided by the MEGaN simulation, a high-resolution direct N-body model of an entire galactic nucleus containing 42 GCs orbiting a SMBH \citep{ASCD17}.

Our main results can be summarized as follows:
{

\begin{itemize}

\item[i)] using the high-resolution data of the MEGaN simulation \citep{ASCD17}, we inferred the amount of BHs delivered by the infalling GCs and captured by the SMBH. We found that the increase in the stellar density around the SMBH causes a corresponding increase of the EMRIs coalescence rate by a factor of 2. Our calculations suggest a  rate $\Gamma_{\rm EMRI} = 0.02-0.17$ yr$^{-1}$ Gpc$^{-3}$. We note that GC infall process acts as a ``segregation booster'', increasing the compact remnants number density around the SMBH on the GCs dynamical friction time-scale;

\item[ii)] assuming that the 2 most massive GCs in our model deliver to the galactic centre two IMBHs, we simulated the evolution of a triple system composed of an IMBH-SMBH binary system orbited by an outer IMBH. We found that the probability for an IMBH-SMBH merging to occur is $51\%$. In $75\%$ of these mergers, the inner binary undergoes a component swap before the merger, thus implying that the events involve the initially outer IMBH. The corresponding rate for SMBH-IMBH merging calculated at redshift $z=0$ is $\Gamma_{\rm SMBH-IMBH} = 0.03$ yr$^{-1}$ Gpc$^{-3}$;

\item[iii)] in the assumption that IMBH formation is particularly efficient, we modelled the evolution of 7 IMBHs orbiting around the SMBH. We grouped simulations in 4 different groups, where we alternatively turned on/off the dynamical friction term and we assumed either a Schwarzschild or Kerr SMBH. We found that 1-3 IMBHs merge with the SMBH within a Hubble time, quite independently of the group considered; 

\item[iv)] after an SMBH-IMBH merger, the resulting GW kick experienced by the remnant can significantly affect its orbital motion if it has a mass $\sim 10^5-10^6\Ms$ and if it is hosted at centre of a dwarf galaxy. We estimate a number density for dwarf galaxies that witnessed an IMBH-SMBH merger and subsequent recoil to be $n_{\rm recoil} \sim 10^5$ Gpc$^{-3}$; 

\item[v)] the complex dynamical interactions between these massive objects lead to the ejection of several IMBH in almost all the cases investigated. The IMBHs reach velocities up to $10^3$ km s$^{-1}$, reaching distances up to $\sim 1-10$ kpc in less than 10 Gyr;

\item[vi)] we investigated whether a stellar BHB can long-live around the IMBH after the disruption of their parent GC. We found that in a few cases the BHB merges before the triple BHB-IMBH system get sufficiently close to the galactic centre. This corresponds to a BHB merger rate of the order of $\Gamma_{\rm BHB,IMBH} = 2$ yr$^{-1}$ Gpc$^{-3}$, comparable to the rate expected for NSCs in galaxies without an SMBH;

\item[vii)] in some cases, the BHB disrupts and one of the components binds to the IMBH, forming a binary system that eventually merge within a Hubble time. The IMBH-BH system formed this way represents the prototype for IMRIs, and is characterised by a merger rate $\Gamma_{\rm BH-IMBH} = 9.5$ yr$^{-1}$ Gpc$^{-3}$;

\item[viii)] we investigate the case in which the IMBH formation fails and GCs leave around the SMBH only several BHBs. In this case, KL oscillations drive the BHB to coalescence over a time-scale of $\sim 2-3$ Myr. This extremely fast channel for BHB coalescence is characterised by a merger rate  $\Gamma_{\rm BHB,SMBH} = 1$ yr$^{-1}$ Gpc$^{-3}$;

\item[ix)] due to the rapid motion around the SMBH, merging BHBs can emit GWs whose frequency is Doppler shifted. Due to this, merging occurring in the densest galactic regions can `appear' up to $30\%$ heavier to detectors. This is an important issue to be taken into account in interpreting the observational results.

\end{itemize}
}

\section*{Acknowledgements}
{ 
The authors acknowledge the anonymous referee for his comments and suggestions that helped us to improve the results presented in this work.
The authors acknowledge Abraham Loeb, Daniel D'Orazio and Monica Colpi, who provided useful comments and suggestions that helped in improving the manuscript. The authors thanks Martina Donnari for having assisted us in the use of the Illustris public data, and Seppo Mikkola, who provided his AR-CHAIN code and assisted us in handling its usage.
MAS acknowledges Sapienza, University of Rome, which funded the research program ``MEGaN: modelling the evolution of galactic nuclei'' via the grant 52/2015 and the Sonderforschungsbereich SFB 881 "The Milky Way System" (subproject Z2) of the German Research Foundation (DFG) for the financial support provided. This work benefited of financial support from the Alexander von Humboldt Foundation, which granted MAS research program ``Chasing black hole at all the scales''. 
This work benefited from support by the ISSI (Bern), through its Intern. Team prog. ref. no. 393 {\it The Evolution of Rich Stellar Populations \& BH Binaries} (2017-18).
Most of the numerical simulations of this paper have been performed on the hybrid CPU+GPU platforms of the ASTRO group of the Dep. of Physics (Sapienza, university of Roma). A part of the numerical simulations presented here were performed on the Milky Way supercomputer, which is funded by the Deutsche Forschungsgemeinschaft (DFG) through the Collaborative Research Center (SFB 881) "The Milky Way System" (subproject Z2) and hosted and co-funded by the J\"ulich Supercomputing Center (JSC).
}

\footnotesize{
\bibliographystyle{mn2e}
\bibliography{ASetal2015}
}

\end{document}